\documentstyle[aps,preprint,epsfig]{revtex}

\tighten

\newcommand{\dx}{\mbox{\rm d}}

\newcommand{\tb}{ \tan \beta}

\newcommand{\mrm}[1]{\;\mbox{\rm #1}}
\newcommand{\beq}{\begin{equation}}
\newcommand{\eeq}{\end{equation}}
\newcommand{\nn}{\nonumber}
\newcommand{\bea}{\begin{eqnarray}}
\newcommand{\eea}{\end{eqnarray}}

\newcommand{\rfn}[1]{(\ref{#1})}
\newcommand{\Eq}[1]{Eq.(\ref{#1})}

\newcommand{\ea}{{\it et al.}}

\newcommand{\eg}{{\it e.g. }}

\newcommand{\np}[1]{Nucl. Phys. {\bf #1}}
\newcommand{\plt}[1]{Phys. Lett. {\bf #1}}
\newcommand{\pr}[1]{Phys. Rev. {\bf #1}}
\newcommand{\prlt}[1]{Phys. Rev. Lett. {\bf #1}}
\newcommand{\zp}[1]{Z. Phys. {\bf #1}}
\newcommand{\epj}[1]{Eur. Phys. J. {\bf #1}}
\newcommand{\prep}[1]{Phys. Rep. {\bf #1}}

\def\lsim{\mathrel{\vcenter{\hbox{$<$}\nointerlineskip\hbox{$\sim$}}}}
\def\gsim{\mathrel{\vcenter{\hbox{$>$}\nointerlineskip\hbox{$\sim$}}}}

\begin{document}

\preprint{\vbox{\baselineskip=13pt
\rightline{UCRHEP-T276 }
\rightline{HIP-2000-23/TH}
\rightline{MZ-TH/00-20}
\vspace*{1cm}}}


\title{Flavour Violation in SUSY SU(5) GUT at Large tan$\beta$}

\author{G. Barenboim$^a$, K. Huitu$^b$, and M. Raidal$^{c,d}$}

\address{\vspace*{0.5cm}
{\small 
$^a$ Institut f\"ur Physik, Johannes Gutenberg Universit\"at, D-55099,
Mainz, Germany \\
$^b$ Helsinki Institute of Physics, P.O.Box 9, FIN-00014 University of
Helsinki, 
Finland\\
$^c$ Department of Physics, University of California, Riverside,
CA 92521, U.S.A. \\
$^d$ National Institute of Chemical Physics, 
R\"avala 10, 10143 Tallinn, Estonia}}


\maketitle 

\vspace*{1cm}

\begin{abstract}
We study flavour violation in the minimal SUSY SU(5) GUT assuming
all the third generation Yukawa couplings to be due to the renormalizable
physics above GUT scale. At large $\tan\beta,$ as suggested by Yukawa
unification in SU(5), sizable flavour violation in the left (right) slepton 
(down squark) sector is induced due to renormalization effects
of down type Yukawa couplings 
between GUT and Planck scales in addition to the flavour violation
in the right slepton sector. The new flavour physics contribution
to $K-\bar K,$ $B-\bar B$ mixing is  small but might
be of phenomenological interest in the case of $b\to s\gamma.$ 
The sign of the latter contribution is the same as the sign of the 
dominant chargino contribution, thus making the constraints on
SUSY scale coming from $b\to s\gamma$ somewhat more restrictive.
The most important feature of the considered scenario is the
large rate of lepton flavour violation. Given the present experimental
constraints, the $\mu\to e\gamma$ and $\mu-e$ conversion branching 
ratios are above the sensitivity of the planned experiments unless the 
SUSY scale is pushed above one TeV.
\end{abstract}

\newpage 


\section{Introduction}

The minimal supersymmetric standard model (MSSM) \cite{NHK,spm}, 
one of the best
motivated extensions of the Standard Model (SM), has triggered 
intensive  research both in theoretical as well as in experimental
physics. Despite of non-trivial constraints on its parameter space 
coming from collider and low energy experiments the MSSM has so far
successfully passed all the tests of precision physics.

In the most general case MSSM contains more than one hundred free
parameters. Some of them may give rise to unobserved phenomena like 
proton decay, large electron dipole moments, large flavour violation,
etc. To explain the absence of such phenomena additional assumptions are
needed to explain the pattern of supersymmetry (SUSY) breaking
parameters. 
An attempt towards that direction is to regard the MSSM as 
a low energy remnant of some grand unified theory (GUT) such as
SU(5), which predicts unification of gauge couplings.
Since the unification naturally occurs in the MSSM, one might try to apply
a similar organizing principle to the soft SUSY breaking masses.
Another prediction of GUTs is the unification of some or all
Yukawa couplings of each generation. 
To achieve successful Yukawa unification \cite{yuk,hall,pol,bdqt,CELW}
the ratio between the vacuum expectation values (vevs) of the two MSSM
Higgs doublets, $\tan\beta,$ is found to be in the range
 $\tan\beta\sim 30 - 50$ for  tau-bottom unification and 
$\tan\beta\sim 50$ for tau-bottom-top unification \cite{bdqt}.  
However,  large mixings in the lepton sector may 
somewhat change this picture \cite{CELW}.

Stringent tests of SUSY GUTs are offered by flavour violation
experiments. The flavour violation in SUSY theories 
is not suppressed by the high scale where the SUSY breaking 
parameters are generated but rather by the mass scale of 
these terms themselves \cite{hkr} which is 
believed to be of order TeV. 
In the scheme of the minimal
flavour violation of Barbieri and Hall \cite{bh} the SUSY breaking
parameters are generated above the GUT scale $M_{GUT}\sim 2\times 10 ^{16}$
GeV, at the reduced Planck scale $M_P\sim 2.4\times 10^{18}$ GeV
by gravitational interactions and are therefore universal at $M_P.$
The renormalization group (RG) evolution below $M_P$ induces 
non-universalities of the soft terms at the GUT scale where the 
SU(5) gauge group  breaks down to the usual MSSM gauge symmetry group. 
Therefore the flavour mixings present in the 
Yukawa couplings at GUT scale cannot be rotated away and should be
reflected at low energies in the squark and slepton mass matrices.
The phenomenology of this scenario in SU(5) SUSY GUT has been studied 
with the assumption that the top quark Yukawa coupling is the only 
sizable one \cite{bhs} which implies flavour 
violation in the right-handed slepton and left-handed down squark sectors. 
Rates of the lepton flavour violating (LFV) processes $\mu\to e\gamma,$
$\mu-e$ conversion in nuclei and $\tau\to \mu\gamma$ are found to be
large for some parts of the SUSY 
parameter space but due to the cancellations 
between gaugino and higgsino loops, they  almost vanish for some other parts 
of the parameter space \cite{japs}.

Another set of SUSY theories predicting large rates of flavour violation
are the ones with right-handed massive neutrinos \cite{CELW,japsN}. 
These models are motivated by the Super-Kamiokande results which
imply maximal mixing between tau and muon neutrinos.    
The large mixing in the neutrino sector induces large flavour mixings
in the left-handed slepton and right-handed down squark sectors in these
models. The rates of flavour violating processes may in this case be much
larger than in the minimal model, for example, the branching ratios of 
LFV processes can be close to or even exceed the 
present experimental bounds.

The aim of the present work is to revisit the minimal flavour violation 
scenario in the SUSY SU(5) grand unified theory.
We extend the considerations of the previous papers \cite{bhs,japs}
in several directions. Following the hints of possible Yukawa 
unification\footnote{Values of $\tb$ between 0.7 and 1.8 are
excluded by the direct LEP2 searches for the lightest 
MSSM Higgs boson \cite{moriond}. At the end of LEP2 run, values 
of  $\tb$ below 2.6
will be probed \cite{dhtw}. This constraint is not very restrictive
but hints into the same direction as the Yukawa unification.} 
we consider the case when all third generation Yukawa couplings
are large and given by the renormalizable physics above the
GUT scale. In this case, sizable flavour violation in the 
left-handed (right-handed) slepton (down squark) sector is induced due to 
the renormalization effects of down type Yukawa couplings 
between GUT and Planck scales, in
addition to the flavour violation in the right-handed slepton sector.
Thus the pattern of flavour violation in SUSY SU(5) GUT at large $\tb$ 
resembles the flavour violation pattern in models with massive 
neutrinos, and deserves phenomenological studies.
We do not make an attempt to identify and to study only that part of 
the soft terms parameter space in which tau-bottom Yukawa unification 
is achieved starting from the low energy parameters, but instead
assume that there may be large corrections depending on the details of
the GUT theory. 
We allow $\tb$ to be a free, although large, parameter
so that the effects of $\tau$ and $b$ Yukawa couplings become
non-negligible. The details of our calculations are given below.
In our numerical analyses we consider the case where  
Yukawa unification is possible as well as the case where it is not.

In addition, we calculate the new flavour physics contributions
to $K-\bar K$ and $B-\bar B$ mixings as well as  to $b\to s\gamma,$ 
$\mu\to e\gamma,$ $\mu-e$ conversion in nuclei and $\tau\to\mu\gamma$
branching ratios. We also comment on the possibility of large 
GUT phases in this context. 
There is an increasing amount of constraints on the MSSM 
mass spectrum coming from direct collider searches as well as from
indirect low energy measurements. We take these  into account
when calculating  allowed ranges for the physical observables. 
In particular, the constraints coming from $b\to s\gamma$ turn out to 
allow only the LFV processes to be in the phenomenologically interesting
ranges.

The main motivation for the present work is to study how sensitively 
the planned next generation LFV experiments will probe the SUSY model
considered here. In the near future the branching ratio of the decay
$\mu\to e\gamma$ will be probed with the sensitivity below $10^{-14}$
\cite{mueprop} and $\mu-e$ conversion in nuclei below $10^{-16}$ \cite{meco}
(in more distant future $\mu$ factories may further reduce these numbers by 
orders of magnitude \cite{ellis}).
We shall show that if these experiments will indeed reach the planned 
sensitivity then, in the scenario considered in this work, LFV must 
be observed unless the SUSY scale $\sqrt{m_{\tilde t_1}m_{\tilde t_2}}$ 
is  above TeV scale.  
This conclusion remains valid even in the case where the only source 
of LFV is the right-handed slepton mass matrix as in Refs. \cite{bhs} 
since the deep cancellation in the branching ratios observed in 
Ref. \cite{japs} occurs at slepton masses which imply multi-TeV SUSY scale.  
In this context it is interesting to note that if we require 
approximate $\tau$-$b$ Yukawa unification at GUT scale then the
sparticle masses are required to be too high for direct production 
at future collider experiments. In this case, the observation of LFV 
may turn out to be the only signal of supersymmetry.

The paper is organized as follows.
In the second section we present the crucial parts of model we are
using throughtout this work, especially the relevant mixing matrices are
introduced.
In section III renormalization from the reduced Planck scale to the
electroweak scale is discussed.
In section IV we present our numerical results.
Finally in section V we summarize our conclusions.

\section{Sources of Flavour Violation}

In this section we present some details of the minimal flavour violation
scenario considered in our work. We assume that the supersymmetric SU(5) 
grand unified theory is valid at mass scales between $M_P$ and $M_{GUT}.$
In the SU(5) model there are three generations of matter multiplets 
$\psi_i$ and $\phi_i,$ $i=1,$ 2, 3, which form the  ${\bf 10}$ and 
${\bf 5}^\ast$ dimensional  representations of SU(5), respectively, and 
${\bf 5}$ and ${\bf 5}^\ast$ dimensional representations of Higgs multiplets, 
$\hat H_2$ and $\hat H_1,$ respectively. 
The tenplets $\psi_i$ contain the quark doublets, the charged lepton singlets
and the up-type quark singlets, while the down-type quark singlets and
the lepton doublets are included in the fiveplets $\phi_i$.
The Higgs fiveplet $\hat H_1$ contains the MSSM Higgs multiplet $H_1$ 
and a coloured Higgs multiplet $H_{C1}$, and $\hat H_2$ contains 
the second MSSM Higgs multiplet $H_2$ and another coloured Higgs multiplet 
$H_{C2}$. 
An adjoint representation Higgs multiplet $ \Sigma$ causing the
breaking of SU(5) should also belong to the Higgs sector of the model. 
We neglect the Yukawa coupling and the soft 
SUSY-breaking parameters associated with it. Inclusion of these 
terms will {\it increase} the non-universalities at GUT scale and thus
{\it increase} the amount of flavour violation in the model. 
Thus the terms in superpotential $W$  relevant for our consideration 
are given by
\begin{eqnarray}
W&=&\frac14 f_{u_{ij}} \psi_i^{AB} \psi_j^{CD} \hat H_2^E \epsilon_{ABCDE} 
  + \sqrt{2} f_{d_{ij}} \psi_i^{AB} \phi_{jA} \hat H_{1B}  ,
\label{su5w}
\end{eqnarray}
where $A,B,...=1,...,5$ are the SU(5) indices.
The relevant soft SUSY breaking terms associated with
the SU(5) multiplets are
\begin{eqnarray}
\label{su5soft}
-{\cal L}_{\rm SUSY~breaking} 
&=&
 (m_{10}^2)_{ij}  \tilde{\psi}_i^{\dagger} \tilde{\psi}_j
+(m_{5}^2)_{ij}  \tilde{\phi}_i^{\dagger} \tilde{\phi}_{j} 
+m_{ h_1}^2  h_1^\dagger  h_1 +  
m_{ h_2}^2 {h}_2^\dagger {h}_2 
\nonumber\\
&&
+\left\{
 \frac14  A_{u_{ij}} \tilde{\psi}_i \tilde{\psi}_j  h_2
+ \sqrt{2}A_{d_{ij}} \tilde{\psi}_i \tilde{\phi}_{j} {h}_1
 +  h.c.
\right\},
\end{eqnarray} 
where $\tilde{\psi}_i$ and  $\tilde{\phi}_i$
 are the scalar components of the ${\psi}_i$ and
${\phi}_i$ chiral multiplets, respectively, and 
$ h_1$ and ${h}_2$ are the Higgs multiplets. 

In the minimal SUGRA scenario the soft SUSY breaking parameters
are generated by gravitational interactions and are universal at $M_P$: 
\begin{eqnarray}
&(m_{10}^2)_{ij}=(m_{5}^2)_{ij}=\delta_{ij} m_0^2,&
\nonumber\\
&m_{ h_1}^2 = m_{{h}_2}^2= m_0^2,&
\nonumber\\
& A_{u_{ij}}=(A'_u\cdot f_{u})_{{ij}}, \,  
A_{d_{ij}}=(A'_d\cdot f_{d})_{{ij}}, \, &
\nn\\
& A'_{u_{ij}}=A'_{d_{ij}}=\delta_{ij} A_0. &
\label{su5init}
\end{eqnarray}
Below $M_P$ till $M_{GUT}$ the parameters in \Eq{su5w} and 
\Eq{su5soft} evolve with energy according to the renormalization 
group equations (RGE) of SUSY SU(5). These can be found for example 
in Ref. \cite{pp} and we do not present them here.
Because the third generation Yukawa couplings $f_{u_{33}}$ and $f_{d_{33}}$
are much larger than the Yukawa couplings of the first two generations,
large hierarchies in the parameters of \Eq{su5init} are induced at 
$M_{GUT}.$ Therefore the soft mass terms $m_{10}^2,$ 
$m_{5}^2$ and $A'_{u,d}$ remain diagonal in the 
generation space but the third generation masses are smaller than
the masses of the first two generations, which remain degenerate
to a good approximation.

To understand the origin of flavour violation in SUSY SU(5) GUT
we have to discuss the evolution of the Yukawa coupling matrices 
$f_d$ and $f_u$  with energy. At $M_P$ we can choose the basis in 
which  $f_u$ is diagonal, $f^P_u=(f^P_u)_{ii}\delta_{ij}.$  
In that case the Yukawa matrix $f_d$ can be diagonalized with a 
bi-unitary transformation, 
\bea
V^{P\dagger} \Theta^P f^P_d U^P,
\eea
where $V^P$ corresponds to the Cabibbo-Kobayashi-Maskawa (CKM) 
matrix at Planck scale and 
$\Theta^P=\mrm{diag}(e^{i\phi_1},e^{i\phi_2}, e^{i\phi_3}),$ 
where $\phi_{1,2,3}$ are the GUT phases (satisfying $\phi_1+\phi_2+\phi_3=0$) 
present in addition to the phase in  $V^P.$ 
However, the mixing matrix $U^P$ can be rotated away by a redefinition of 
the fiveplet fields $\phi.$ No trace of that rotation will remain
in the soft SUSY breaking terms because these are universal and
proportional to unit matrix at $M_P.$ Below $M_P$
the  Yukawa matrices $f_d$ and $f_u$ run according to 
\bea
16\pi^2\frac{\dx}{\dx t} f_{u_{ij}} &=& 
\left[ -\frac{96}{5} g_5^2 
       + 3{\rm Tr}(f_u^\dagger f_u) 
\right] f_{u_{ij}} 
+ 6 (f_u f_u^\dagger f_u)_{ij}
+ 2 (f_d f_d^\dagger f_u)_{ij} 
+ 2 (f_u f_d^\ast f_d^\top)_{ij} , \nn \\
16\pi^2\frac{\dx}{\dx t} f_{d_{ij}} &=& 
\left[ -\frac{84}{5} g_5^2 + 4{\rm Tr}(f_d^\dagger f_d) \right] f_{d_{ij}}
+ 6 (f_d f_d^\dagger f_d)_{ij} 
+ 3 (f_u f_u^\dagger f_d)_{ij}   .
\label{su5yuk}
\eea
Because the third generation Yukawa couplings are large, of order unity,
the running of the off-diagonal elements of the Yukawa matrices is 
significant. Notice that despite of the chosen basis,  non-zero 
off-diagonal elements are generated also in $f_u.$ To predict
the running of the off-diagonal elements requires a precise knowledge
of all the elements of the Yukawa matrices. This, however, goes
beyond the assumptions made in our work: we assumed that only  the
third generation Yukawa couplings can be reliably estimated via the RGEs
from the experimental data. Therefore, in our numerical estimates 
we assume that the off-diagonal elements of the rotation matrix $U$,
which rotates the fiveplet fields $\phi$ are small, of order 
or smaller than the corresponding CKM matrix elements at GUT scale.

At GUT scale  the Yukawa coupling matrices  $f_d$ and $f_u$ get
renormalized to $f^G_d$ and $f^G_u,$ respectively, and can be 
diagonalized again with bi-unitary transformations.
However, now any rotation of the superfields $\psi$ and $\phi$ 
is reflected in the soft SUSY breaking  parameters because 
these are not universal at GUT scale any more. These rotations give rise
to large flavour changing effects, as will be discussed in the following.

At $M_{GUT}$ the SU(5) gauge group breaks spontaneously into the
usual MSSM gauge group $\mrm{SU(3)}_C\times\mrm{SU(2)}_L\times\mrm{U(1)}_Y.$
The MSSM superpotential $W$ valid below GUT scale is 
\begin{eqnarray}
\label{mssmw}
W&=&  Q^o_i (f_{u_{ij}}) U^{co}_i H_2
  -  Q^o_i (f_{d_{ij}}) D^{co}_{j} H_1 
  -  E^{co}_i (f_{e_{ij}}) L^o_j H_1 - \mu H_1 H_2 \,
\end{eqnarray}
and the soft SUSY breaking terms are given by
\begin{eqnarray}
-{\cal L}_{\rm soft}&=& 
  \tilde{L}_{i}^{o\dagger} (m_{\tilde L}^2)_{ij}\tilde{L}^o_{j} 
+ \tilde{ E}_{i}^{co\ast} (m_{\tilde E}^2)_{ij}\tilde{E}_{j}^{co}  
+ \tilde{Q}_{i}^{o\dagger}(m_{\tilde Q}^2)_{ij} \tilde{Q}^o_{j} 
+ \tilde{U}_{i}^{co\ast} (m_{\tilde U}^2)_{ij} \tilde{U}^{co}_{j}  
+ \tilde{D}_{i}^{co\ast} (m_{\tilde D}^2)_{ij} \tilde{D}^{co}_{j}  
\nonumber\\
&& +m_{H_1}^2 H_1^\dagger H_1  
   +m_{H_2}^2 H_2^\dagger H_2  
 + \left(\tilde{Q}^o_{i} (A_{u_{ij}})\tilde{U}^{co}_{j} H_2
  -\tilde{Q}^o_{i} (A_{d_{ij}})\tilde{D}^{co}_{j}  H_1
  -\tilde{E}^{co}_{i} (A_{e_{ij}})\tilde{L}^o_{j} H_1 \right. \nn\\
&&
    \left. +B_{H} H_1H_2 
 +\frac12 M_1 \tilde{B} \tilde{B}
 + \frac12 M_2 \tilde{W}^a \tilde{W}^a 
 + \frac12 M_3 \tilde{g}^a \tilde{g}^a
   + h.c.\right) .
\end{eqnarray}
Here all the fields are explicitly written in the flavour eigenstate 
basis denoted by the superscript zero. 
The boundary condition for the soft terms are specified via
\bea
& (m^2_{\tilde Q})_{ij}=(m^2_{\tilde U})_{ij}=(m^2_{\tilde E})_{ij}=
(m^2_{10})_{ij}\,, & \nn\\
& (m^2_{\tilde D})_{ij}=(m^2_{\tilde L})_{ij}=(m^2_5)_{ij} \,,& \nn\\
& m^2_{H_1}=m^2_{ h_1}\,, \;\;\; m^2_{H_2}=m^2_{ h_2}\,, & \nn\\
& M_1=M_2=M_3=M_0.&  
\label{mssmbound}
\eea
Since there is no splitting among the gaugino masses above the 
GUT scale, it is most convenient to stipulate $M_0$ at the GUT scale. 
For the Yukawa couplings and trilinear terms this model predicts
\bea
f^G_e=f^G_d\,,\;\;\; A^G_e=A^G_d\,
\eea
at GUT scale.

From the GUT scale down to the electroweak scale the Yukawa couplings and 
the soft parameters evolve with energy via the MSSM RGEs \cite{MV}.
Once the electroweak symmetry is broken, rotations of the superfields
\bea
\begin{array}{ccc}
D^o=V_dD\,,\; &  E^o=V_eE\,,\; &  U^o=V_uU\,,  \\
D^{co}=U^*_dD^c\,, &  E^{co}=U^*_eE^c\,, &  U^{co}=U^*_uU^c\,, 
\end{array}
\label{rot}
\eea
bring quarks and leptons into their mass eigenstates with diagonal 
Yukawa couplings $f'_d$, $f'_e$ and $f'_u$
\bea
f'_{d_i}=(V^T_d f_d U_d^*)_{ii}\,,\;\;
f'_{e_i}=(U_e^\dagger f_e V_e)_{ii}\,,\;\;
f'_{u_i}=(V^T_u f_u U_u^*)_{ii} \,.
\eea
At the GUT scale the down quark and lepton masses are predicted to be
equal. Therefore the diagonalizing matrices are related at $M_{GUT}$
as
\bea
V_d^G=U_e^{*G}\,,\;\;\; U_d^G=V_e^{*G}.
\label{vgut}
\eea
This implies that at low energies the left-handed quark rotation 
matrix, the CKM matrix in the basis in which up quark Yukawa matrix 
is diagonal, can be related to the right-handed lepton rotation matrix
and the right-handed down quark rotation matrix can be related 
to the left-handed lepton rotations via the RGEs.

At the same time the rotation in \Eq{rot} changes the basis of the 
superpartners of quarks and leptons. For example,
the mass eigenstates of the charged sleptons and sneutrinos
can be  expressed as
\bea
\tilde E=\Gamma_E\left(
\begin{array}{c}
V_e^\dagger\tilde E^o  \\
U_e^\dagger\tilde E^{co\ast}
\end{array}
\right)\,,\;\;
\tilde \nu=\Gamma_\nu V_e^\dagger \tilde \nu^o\,,
\eea
where $\Gamma_E$ and $\Gamma_\nu$ are $6\times 6$ and $3\times 3$ 
rotation matrices,
respectively. The slepton mass  matrices  are given by
\bea
 m^2_{\tilde E} &=& \Gamma_E \left(
\begin{array}{cc}
V_e^\dagger m_{\tilde L}^2 V_e + m_{e}^2 
- m_Z^2  \cos 2 \beta (\frac12 - \sin^2 \theta_W)  & 
-  \mu m_{e}  \tan\beta + m_e U_e^\dagger A^{'\dagger}_e U_e 
 \\
 & \\
-  \mu^* m_{e} \tan\beta + U_e^\dagger A'_e U_e  m_e  
 &
U_e^\dagger m_{\tilde E}^2  U_e + m_{e}^2  
- m_Z^2  \cos 2 \beta  \sin^2 \theta_W  
\end{array}
\right) \Gamma^{\dagger}_E  \nn \\
&& \nn \\
 m^2_{\tilde \nu}&=& \Gamma_\nu \left( 
V_e^\dagger m_{\tilde L}^2 V_e  + \frac12  m_Z^2  \cos 2 \beta
\right)\Gamma^{\dagger}_\nu  \,,
\label{slmatr}
\eea
and analogously for squarks.

Notice that this definition of the diagonalizing matrices 
$\Gamma_{E,D,U}$
differs from the one originally given in Ref. \cite{bbmr}
where the matrices $\Gamma$ relate the flavour eigenstates directly to
the mass eigenstates. The advantage of the present notation
in calculating the flavour violating observables in our scenario 
is the following.  It allows us to perform the MSSM RGE running of the 
soft terms (which are diagonal in the flavour space) 
and the CKM matrix elements (which induce the flavour mixings) 
separately without constructing the squark mass matrices at high
scales and without running each element of the $6\times 6$ sparticle
mass matrices. 
The slepton and squark mass matrices are then constructed 
at low energies in terms of the low energy values of the 
soft terms, quark masses and the mixing matrices.

We use the standard notation for the neutralino and chargino 
mass matrices. The neutralino mass matrix $M_{\tilde N}$ 
in the basis $(\tilde B, \tilde W^0, \tilde H_1^0,  \tilde H_2^0)$
can be diagonalized as
\bea
M_{\tilde N}^{diag}=N M_{\tilde N} N^\dagger
\eea
where $N_{ij}$ is a $4\times 4$ unitary matrix. Similarly, the 
two diagonalizing matrices $O_{L,R}$ in the chargino sector 
can be found from
\bea
M_{\tilde \chi}^{diag}= O_L 
\left(
\begin{array}{cc}
M_2 & \sqrt{2} \sin\beta m_W \\
\sqrt{2} \cos\beta m_W & \mu
\end{array}
\right) O_R^\dagger
\eea
For the further details see, \eg, Ref. \cite{spm}.

\section{Renormalization Procedure}

Before calculating the rates of flavour violating observables,
let us describe our procedure of calculating  the input SUSY parameters
via the RGE evaluation. We start the RGE running at $M_Z$
where we introduce as the low energy input the values of
the gauge coupling constants, the tau and bottom quark Yukawa
couplings corresponding to the tau and bottom quark masses
$m_\tau(M_Z)=1.784$ GeV and $m_b(M_Z)=3.0$ GeV \cite{rsb}, respectively.
We  parametrize the CKM matrix in the standard way \cite{pdb} with
$\theta_{12}=0.22,$ $\theta_{23}=0.04$ and $\theta_{13}=0.003.$
We evolve these quantities from $M_Z$
to $m_t$ using two loop SM RGE-s for five flavours. At $m_t$
we include top quark Yukawa coupling corresponding to the
pole mass $m_t=174$ GeV and run the SM RGE-s for six flavours up
to the scale $Q$ where superparticles are introduced.
At the scale $Q$ we convert the SM couplings to the MSSM ones fixing
the value of $\tb$ and include the SUSY loop corrections 
to the bottom quark and tau lepton Yukawa 
couplings according to Ref. \cite{hall,guni}. 
Then the gauge and Yukawa
couplings are evaluated at GUT scale with the help of two loop MSSM
RGE-s. For the running of the CKM matrix elements we use
the one loop RGE-s from Ref. \cite{op}.

The crucial issue in the present context is the correct tau-bottom Yukawa
unification as predicted by SU(5). The Yukawa unification has been studied 
extensively in literature \cite{yuk,hall,bdqt} and 
it turns out that unification 
is possible only for $\mu < 0$ and $30\lsim\tb\lsim 50.$ 
This can be understood as follows. The correction to the bottom
Yukawa coupling \cite{hall}, 
\bea
f_b\sim\frac{m_b}{1+\delta_b}\frac{1}{v_1}
\eea
which is numerically the most important one,
is proportional to $\tb$ and its sign depends on $sign(\mu).$     
Tau-bottom Yukawa unification can be obtained if the top 
quark mass is approximately at a fixed point at $\tb\approx 1.6.$
On the other hand, small values of $\tb$ are, anyway, 
excluded by the LEP 2 search for the
lightest Higgs boson \cite{moriond,dhtw}. 
To achieve the tau-bottom Yukawa unification 
therefore  a sizable negative $\delta_b$ is needed, which requires
large values of $\tb$ and fixes the sign of the $\mu$ parameter to be
 $sign(\mu)=-1.$ On the other hand, it has been argued \cite{pol} that
there might be important threshold corrections due to the coloured
triplet Higgses at GUT scale modifying sizably the bottom quark 
Yukawa coupling already at $M_{GUT}.$ In any case, the coupling 
which receives large corrections is the bottom Yukawa coupling,
tau Yukawa coupling is less sensitive to these corrections and thus
the prediction of tau Yukawa coupling at GUT scale is more robust.
Therefore, in our numerical calculations we use actually 
the matching condition 
\bea
f_d^G[\mrm{SU(5)}]=f_\tau^G[\mrm{MSSM}]
\eea 
at $M_{GUT}$ to give a numerical value to the SU(5) Yukawa
coupling $f_{d_{33}},$ and allow the value of the MSSM coupling
$f_b^G$ to be numerically different.
To achieve semi-realistic Yukawa coupling evolution we fix the 
magnitude of $\delta_b$ by hand for each considered value of $\tb$ 
but keep its sign to be determined by
the $sign(\mu).$ In such a case the Yukawa unification is 
achieved for $sign(\mu)=-1$ and not achieved for $sign(\mu)=+1.$
In our numerical examples we consider both possibilities.
For definiteness we consider two values of $\tb,$ $\tb=35$
(corresponds to $f_t^G\sim 2 f_\tau^G$) and $\tb=48$
(corresponds to $f_t^G\sim f_\tau^G$).

At $M_{GUT}$ we fix the leptonic mixing matrices $V_e$ and
$U_e$ according to \Eq{vgut}. 
Unlike the quark mixing matrices  $V_e$ and $U_e$ do not run
with energy due to the absence of neutrino Yukawa couplings in the MSSM.  
Choosing the basis in which the top Yukawa matrix is diagonal
we have $V_d=V_{CKM}$ and therefore $|U_e|_{ij}=|V_{CKM}^G|_{ij}.$  
There is no experimental restriction on the mixing matrix of the
right-handed quark fields. Therefore we assume that large couplings
$f_{d_{33}}, f_{u_{33}}$ in \Eq{su5yuk} generate small non-zero
$(31)$ and $(32)$ elements for $U_d$ and $V_e$ but the angle $\theta_{12}$ 
remains negligible in these matrices. 
Numerically we consider three cases:  $V_e^{(31),(32)}=U_e^{(31),(32)},$
$V_e^{(31),(32)}=0.1\times U_e^{(31),(32)}$ and $V_e^{(31),(32)}=0.$
 
Further, we run the SU(5) gauge and Yukawa couplings from $M_{GUT}$ to
$M_P$ where we fix the values of $m_0$ and $A_0.$ For the trilinear
coupling we take always  $A_0=0$ in our numerical calculations.
Then we run all the parameters, including
the SU(5) soft terms down to the GUT scale, apply the boundary
conditions \Eq{mssmbound} and run all the MSSM parameters 
down to the scale $Q$ which we take to be 200 GeV. At that scale we
fix the SUSY parameters $\mu$ and $B$ via the conditions of 
spontaneous symmetry breaking and calculate the masses and mixing 
matrices of the sparticles. The chargino and neutralino mass
matrices are given by the standard expressions which can be
found for example in Ref. \cite{spm}. 
In order not to run into contradiction with the mass bounds on 
the SUSY particles from direct searches we take $M_0\gsim 150$ GeV
and   $m_0\gsim 150$ GeV; the lightest sparticle masses corresponding
to $m_0=M_0= 150$ GeV and $\tb=48$ are roughly $M_{\tilde \tau_1}=73$ GeV,
$M_{\tilde \chi^+_1}=98$ GeV and $M_{\tilde \chi^0_1}=58$ GeV and
are just on the limit of the present LEP bounds.

\section{Rates of Flavour Changing Processes}

\subsection{$K_L-K_S$ System}

We start our numerical analyses by calculating the new 
SUSY flavour changing
physics contribution to physical observables in the $K_L-K_S$ system.
Recently it has been emphasized \cite{moroi} that in models with massive
neutrinos the SUSY flavour changing contribution to $\epsilon_K$ 
may be large, especially if new GUT phases are present which 
may maximize the effect. Because at large $\tb$ the pattern
of flavour violation in the SUSY SU(5) is the same as in the 
models with right-handed neutrinos we study this issue carefully.

The dominant new physics contribution to $\Delta S=2$ processes
comes from box diagrams with gluinos and down squarks running
in the loop \cite{bbmr}.
Because in our scenario the first two generation sfermions are 
almost exactly degenerate the appropriate tool to use
is the mass insertion approximation \cite{ggms}. The current state 
of the art on this subject is summarized in Ref. \cite{eps}
in which the low energy  $\Delta S=2$ effective Hamiltonian is 
calculated including NLO QCD corrections and the relevant 
hadronic matrix elements are evaluated using the lattice results for 
the $B_K$-parameters. We use the expressions for the  $\Delta S=2$ 
Wilson coefficients, NLO QCD corrections and hadronic matrix elements 
as well as numerical input exactly as in Ref. \cite{eps} and therefore
we do not copy them here. However, we need to specify the model 
dependent input.

In the mass insertion approximation the flavour violation is 
characterized with the dimensionless parameters $(\delta_{LL})_{ij},$
$(\delta_{RR})_{ij},$ $(\delta_{LR})_{ij}$ and $(\delta_{RL})_{ij}$
defined as
\bea
\left(
\begin{array}{cc}
\delta_{LL} & \delta_{LR} \\
\delta_{RL} & \delta_{RR} 
\end{array}
\right) = \frac{1}{\tilde m^2}
\left(
\begin{array}{cc}
(m^2_{\tilde D})_{LL} & (m^2_{\tilde D})_{LR} \\
(m^2_{\tilde D})_{RL} & (m^2_{\tilde D})_{RR} 
\end{array}
\right) 
\eea
where $(m^2_{\tilde D})_{MN},$ $M,N=L,R$ are the corresponding 
$3\times 3$ components of the $6\times 6$ down squark mass matrix and
$\tilde m^2$ is an average squark mass appropriate for the problem
under consideration. We calculate the down squark mass matrix as
described in detail in previous sections. The mixings among the left
down squark fields are generated by the CKM matrix $V_{CKM}.$
The mixings among the right down squarks are assumed to be given by
the matrix $U_{d_{ij}}=(V^G_{CKM})_{ij},$ $i,j=2,3$ with
vanishing angle $\theta_{12}.$  Thus the off-diagonal elements of $U_d$ are
smaller than the elements of $V_{CKM}.$
Because only the $(12)$ components of 
$\delta_{MN}$ enter to the $\Delta S=2$ process we identify
$\tilde m^2=(m^2_{\tilde Q})_{11}.$

\begin{figure}[t]
\centerline{
\epsfxsize = 0.5\textwidth \epsffile{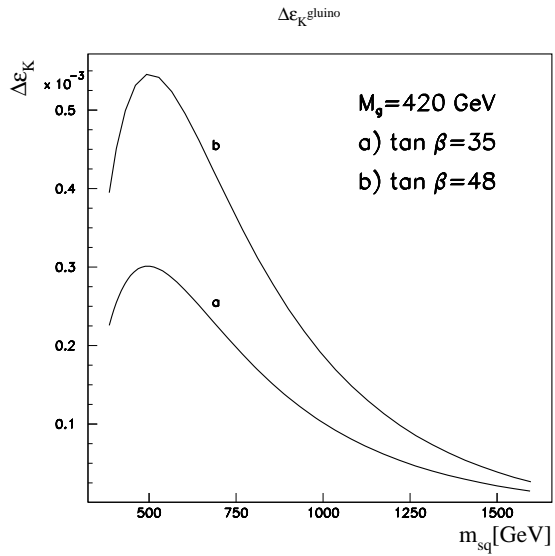} 
}
\caption{\it 
Gluino mediated contribution to $\epsilon_K$ as a function of 
the average down squark mass assuming that the possible 
new GUT phases maximize the effect. 
\vspace*{0.5cm}}
\label{fig:eps}
\end{figure}

The $K_L-K_S$ mass difference and the CP-violating parameter
$\epsilon_K$ are given by:
\bea
\Delta M_K & = & 
2 \mrm{Re} \langle K^0|H_{eff}^{\Delta S=2}| \bar K^0\rangle \,,
\nn\\
\epsilon_K & = & \frac{1}{\sqrt{2}\Delta M_K}
\mrm{Im} \langle K^0|H_{eff}^{\Delta S=2}| \bar K^0\rangle\,.
\eea
It turns out that the new physics contribution  to $\Delta M_K$ 
is very small in our model, 
typically at the permil level of the measured value.
However, the new contribution to $\epsilon_K$ which is a small
quantity in the SM can be sizable. In Fig. \ref{fig:eps}
we plot the gluino mediated contribution to $\epsilon_K$
as a function of the average squark mass $\tilde m$ fixing the 
gluino mass to be $M_{\tilde g}=420$ GeV and assuming that the
new GUT phases maximize the effect. For the squark masses 
of order 500 GeV and large $\tb$ the contribution may exceed
25\% of the measured value
$\epsilon_K=2.3\cdot 10^{-3}.$
However, because of the assumption of the GUT phases and because
of large theoretical errors in the hadronic matrix elements
no useful constraints can be derived from $\epsilon_K$ on the
parameters of the model. The reason why the contribution to 
$\epsilon_K$ in SUSY SU(5) is smaller than in the models with
massive neutrinos is the smallness of the $(\delta_{MN})_{12}$ 
elements. The magnitude of the largest mixing parameter 
$(\delta_{LL})_{12}$ is typically below $10^{-4}$ and for the 
chosen parameters  $(\delta_{RR})_{12}$ is about factor of three smaller.
The (LR) parameters are smaller by three orders of magnitude.
We conclude that the small mixing angles in our approach suppress
the contribution to the $K-\bar K$ mixing.

\subsection{$B_d-\bar B_d$ System}

Here we estimate the new physics contribution to the $B_d$ meson system. 
The SM contribution to $\Delta M_{B_d}=2|M_{12}|$ 
is given by  \cite{buras}
\begin{equation}
M_{12}^{ SM}=\frac{G_{F}^2}{12\pi^2}\eta_{ QCD}B_{B_{d}}f_{B_{d}}^2 
M_{B_{d}}M_W^2(V_{td}V_{tb}^*)^2 S_0(z_t),
\label{smm}
\end{equation}
where 
\begin{equation}
 S_0(z_t)=\frac{4z_t-11z_t^2+z_t^3}{4(1-z_t)^2}
-\frac{3z_t^3\ln z_t}{2(1-z_t)^3}\,,
\end{equation}
where   $z_t=m_t^2/m_W^2$ and $\eta_{ QCD}=0.55\pm0.01$.
With  $B_{B_d}=1.29\pm{0.08}\pm{0.06}$ and $f_{B_d}=175\pm{25}$ MeV
it successfully predicts the measured value  
$\Delta M_{B_d}=0.470\pm 0.019$ ps$^{-1}.$ 
To estimate the magnitude of 
the dominant gluino induced contribution to $B-\bar B$ mixing in our
scenario we use again the mass insertion approximation. 
Neglecting the small LR and RL contributions it reads \cite{ggms} 
\begin{eqnarray}
\lefteqn{M_{12}^{SUSY}=-\frac{\alpha_s^2}{216\tilde m^2}
\frac{1}{3}B_{B_{d}}f_{B_{d}}^2M_{B_{d}}
\left\{((\delta_{LL})_{31}^2+(\delta_{RR})_{31}^2)
(66\tilde{f}_6(x)+24xf_6(x))+\right.}\label{ma}
\\
&&\hspace*{-0.2cm}\left.(\delta_{LL})_{31}(\delta_{RR})_{31}
\left[\left(36-24\left(\frac{M_{B_{d}}}{m_b+m_d}\right)^2\right)
\tilde{f}_6(x)+\left(72+384\left(\frac{M_{B_{d}}}{m_b+m_d}\right)^2
\right)xf_{6}(x)\right]\right\},\nonumber\\ && \nn\\
&&\hspace*{-0.2cm}
f_6(x)=\frac{1}{6(x-1)^5}\left(
x^3-9x^2-9x+17+ 6(1+3x){\rm ln}x \right)\,,
\\
&&\hspace*{-0.2cm}
\tilde{f}_6(x)=\frac{1}{3(x-1)^5}\left(
-x^3-9x^2+9x+1+6x(1+x){\rm ln}x \right) \,.
\end{eqnarray}
where $x=M_{\tilde{g}}^2/{\tilde{m}}^2.$
In this case the common squark mass is taken to be
$\tilde m^2=(m^2_{\tilde Q})_{33}$ and the mixing elements $\delta_{31}$ 
are calculated as in the previous subsection.

In the case where the sparticle masses are
assumed  to take their lower allowed limits, 
the SUSY contribution to $\Delta M_{B_d}$ turns out to be
at most of order a few percent. Taking into account the large 
errors in the input parameter $f^2_{B_d} B_{B_d}$ 
no useful constraints on our model can be derived
drom the measurement of $\Delta M_{B_d}$.

\subsection{$b\to s \gamma$}

The radiative decay $b\to s\gamma$ is known to get large contributions
from SUSY particle loops and therefore this process implies
strong constraints on the allowed SUSY parameter space 
\cite{bbmr,early,msugra}. 
Besides the SM $W$-boson-$t$-quark contribution there are also 
charged Higgs, chargino, neutralino and gluino contributions.
At the electroweak scale all of them have been calculated in Ref. \cite{bbmr}.
While in the SM the NLO QCD corrections are included to  $b\to s\gamma$,
in SUSY theories the NLO analyses has been performed only for 
specific scenarios \cite{qcd}. The complication here lies in the 
new flavour structure of SUSY theories, as  compared to the SM,
which extends the operator basis beyond the SM one. Therefore
the LO QCD corrections to the gluino contribution which gives 
the dominant new flavour physics contribution 
was calculated very recently in Ref. \cite{bghw}. Before that, the 
gluino contribution in a generic SUSY flavour model was
studied in \cite{ggms} without including QCD corrections.

Here we shortly review the results of Ref. \cite{bghw}.
The effective Hamiltonian for $b\to s\gamma$ is expressed in two parts
\bea
{\cal H}_{eff}= {\cal H}_{eff}^{CKM} + {\cal H}_{eff}^{\tilde g}\,,
\eea
where ${\cal H}_{eff}^{CKM}$ is the effective Hamiltonian in
which the structure of flavour violation is the same as in the SM
and the gluino contribution ${\cal H}_{eff}^{\tilde g}$ 
exhibits
the new flavour structure. The Wilson coefficients of the first term,
\begin{equation}
 {\cal H}_{eff}^{CKM} = 
 - \frac{4 G_F}{\sqrt{2}} V_{tb}^{\phantom{\ast}} V_{ts}^\ast
  \sum_i C_i(\mu) {\cal O}_i(\mu) \,, 
\label{weffham}
\end{equation}
contain the SM as well as the charged Higgs, chargino and neutralino
contributions. The dominant operators in \Eq{weffham} 
are the dimension six  magnetic operators (${\overline m}_b(\mu)$
is the running mass)
\bea
{\cal O}_{7}=
  \displaystyle{\frac{e}{16\pi^2}} \,{\overline m}_b(\mu) \,
 (\bar{s} \sigma^{\mu\nu} P_R b) \, F_{\mu\nu}\,,  \;\;\;\;\;\;
{\cal O}_{8} =
  \displaystyle{\frac{g_s}{16\pi^2}} \,{\overline m}_b(\mu) \,
 (\bar{s} \sigma^{\mu\nu} T^a P_R b)
     \, G^a_{\mu\nu}\,,                               
\label{o7}                                        
\eea
and  the operators ${\cal O}'_{7,8}$ obtained by $L\leftrightarrow R.$
The gluino effective hamiltonian
\begin{equation}
 {\cal H}_{eff}^{\tilde{g}} = 
 \sum_i C_{i,\tilde{g}}(\mu) {\cal O}_{i,\tilde{g}}(\mu)  +
 \sum_i \sum_q C_{i,\tilde{g}}^q(\mu) {\cal O}_{i,\tilde{g}}^q(\mu) \,. 
\label{geffham}
\end{equation}
contains in addition to the dimension six magnetic operators 
${\cal O}_{7b,\tilde{g}},$ ${\cal O}'_{7b,\tilde{g}},$ 
${\cal O}_{8b,\tilde{g}},$ ${\cal O}'_{8b,\tilde{g}}$ 
also  operators of dimension five,
\bea          
{\cal O}_{7\tilde{g},\tilde{g}}  = 
  e \,g_s^2(\mu) \,
 (\bar{s} \sigma^{\mu\nu} P_R b) \, F_{\mu\nu}\,,    \;\;\;\;\;\; 
{\cal O}_{8\tilde{g},\tilde{g}}   = 
 g_s(\mu) \,g_s^2(\mu) \,
 (\bar{s} \sigma^{\mu\nu} T^a P_R b)
     \, G^a_{\mu\nu}\,, 
\label{gmagnopg}                                     
\eea
in which the chirality-violating parameter is the 
gluino mass in the corresponding Wilson coefficients, 
as well as  new four-quark operators ${\cal O}^q_{i,\tilde g},$
$q=u,\,d,\,c,\,s,\,b,$  which are listed in Ref. \cite{bghw}.
These new operators change the structure of the LO QCD corrections
in general SUSY flavour models.

The Wilson coefficients in \Eq{weffham} including  the LO QCD corrections 
in the MSSM are well known. Their explicit expressions 
 can be found in \cite{msugra} and we do not rewrite them here.
Instead, let us for a moment concentrate on the study of 
the effective hamiltonian \Eq{geffham} in our model.

At the electroweak scale the relevant Wilson coefficients are
given by \cite{bghw}
\begin{eqnarray}
C_{7b,\tilde{g}}(\mu_W)              & = &
\ \ 
-\frac{e_d}{16 \pi^2} \ C(R)
 \sum_{k=1} ^6 \frac{1}{m_{\tilde{d}_k}^2} 
\left( \Gamma_{DL}^{kb} \, \Gamma_{DL}^{\ast\,ks} \right)
 f_2(x_{gd_k}) \,,
                                     \nonumber \\
C_{7\tilde{g},\tilde{g}}(\mu_W)      & = & 
 M_{\tilde g}\,
 \frac{e_d}{16 \pi^2} \ C(R) 
 \sum_{k=1} ^6 \frac{1}{m_{\tilde{d}_k}^2} 
\left( \Gamma_{DR}^{kb} \, \Gamma_{DL}^{\ast\,ks} \right)
 f_4(x_{gd_k})\,, 
\nn \\
C_{8b,\tilde{g}}(\mu_W)              & = &
\ \ 
-\frac{1}{16 \pi^2} 
 \sum_{k=1} ^6 \frac{1}{m_{\tilde{d}_k}^2} 
\left( \Gamma_{DL}^{kb} \, \Gamma_{DL}^{\ast\,ks} \right) \,  
\left[
\left(C(R) -\!{\scriptstyle{1\over 2}} C(G)\right) f_2(x_{gd_k})
      -{\scriptstyle{1\over 2}} C(G) f_1(x_{gd_k}) 
\right]\,,
                                     \nonumber \\
C_{8\tilde{g},\tilde{g}}(\mu_W)      & = & 
 M_{\tilde g}\, 
 \frac{1}{16 \pi^2} 
 \sum_{k=1} ^6 \frac{1}{m_{\tilde{d}_k}^2} 
\left( \Gamma_{DR}^{kb} \, \Gamma_{DL}^{\ast\,ks} \right) \,            
\left[
\left(C(R) -\!{\scriptstyle{1\over 2}} C(G)\right) f_4(x_{gd_k})
      -{\scriptstyle{1\over 2}} C(G) f_3(x_{gd_k}) 
\right] \,,
\label{glgl}                
\end{eqnarray}
Here the matrices $ \Gamma_{DL}$ and $ \Gamma_{DR}$ are the $6\times 3$
submatrices of $ \Gamma_{D},$
\bea
\Gamma_{D}^{6\times 6}=\left(
\begin{array}{cc}
\Gamma_{DL}^{6\times 3} & \Gamma_{DR}^{6\times 3}\,,
\end{array}
\right)
\eea
and the ratios $x_{gd_k}$ are defined as 
$x_{gd_k} \equiv M_{\tilde g}^2/m_{\tilde{d}_k}^2.$ The Casimir
factors $C(R)$ and $C(G)$ are  $C(R)=4/3$ and $C(G)= 3$
and the functions $f_i(x)$, $i = 1,...,4$, are given by 
\begin{eqnarray}
 f_1(x) & =  &  \frac{1}{ 12\, (\!x-1)^4}
  \left( x^3 -6x^2 +3x +2 +6x\log x\right)  \,,
  \nonumber \\
 f_2(x) &  =  & \frac{1}{ 12\, (\!x-1)^4} 
  \left(2x^3 +3x^2 -6x +1 -6x^2\log x\right) \,,
  \nonumber \\
 f_3(x) &  =  & \frac{1}{\phantom{1} 2\, (\!x-1)^3} 
  \left( x^2 -4x +3 +2\log x\right) \,,
  \nonumber \\
 f_4(x) &  =  & \frac{1}{ \phantom{1} 2\, (\!x-1)^3}
  \left( x^2 -1 -2x\log x\right)\,.
\label{loopfunc}
\end{eqnarray}
The Wilson coefficients of the
corresponding primed operators are obtained through the interchange
$\Gamma_{DR}^{ij} \leftrightarrow \Gamma_{DL}^{ij}.$
At the low scale $\mu_b$, the LO renormalization group improved  
coefficients become
\begin{eqnarray}
{C_{7\tilde{g},\tilde{g}}(\mu_b)}   & = &  
 \eta^{\frac {27}{23}}
 \,{C_{7\tilde{g},\tilde{g}}}(\mu_W) + 
 {\displaystyle \frac {8}{3}} 
 \left( \eta^{\frac {25}{23}} -\eta^{\frac {27}{23}} \right)
 \,{C_{8\tilde{g},\tilde{g}}}(\mu_W)\,,   
\nonumber \\
{C_{7b,\tilde{g}}(\mu_b)}           & = &  
  \eta^{\frac {39}{23}} 
 \,{C_{7b,\tilde{g}}}(\mu_W) + 
{\displaystyle \frac {8}{3}}  
\left( \eta^{\frac {37}{23}} -\eta^{\frac {39}{23}}\right)
 \,{C_{8b,\tilde{g}}}(\mu_W) + {R_{7b,\tilde{g}}}(\mu_b)\,, 
\label{evoldimfive}
\end{eqnarray} 
where  $\eta=\alpha_s(\mu_W)/\alpha_s(\mu_b)$. 
The remainder function $R_{7b,\tilde{g}}(\mu_b)$  turns out to be
numerically very small \cite{bghw} and we neglect it in our numerical 
computation.
The low-scale Wilson coefficients 
for the corresponding primed operators are obtained by  replacing 
in~(\ref{evoldimfive}) all the unprimed coefficients with  
primed ones. 

\begin{figure}[t]
\centerline{
\epsfxsize = 0.5\textwidth \epsffile{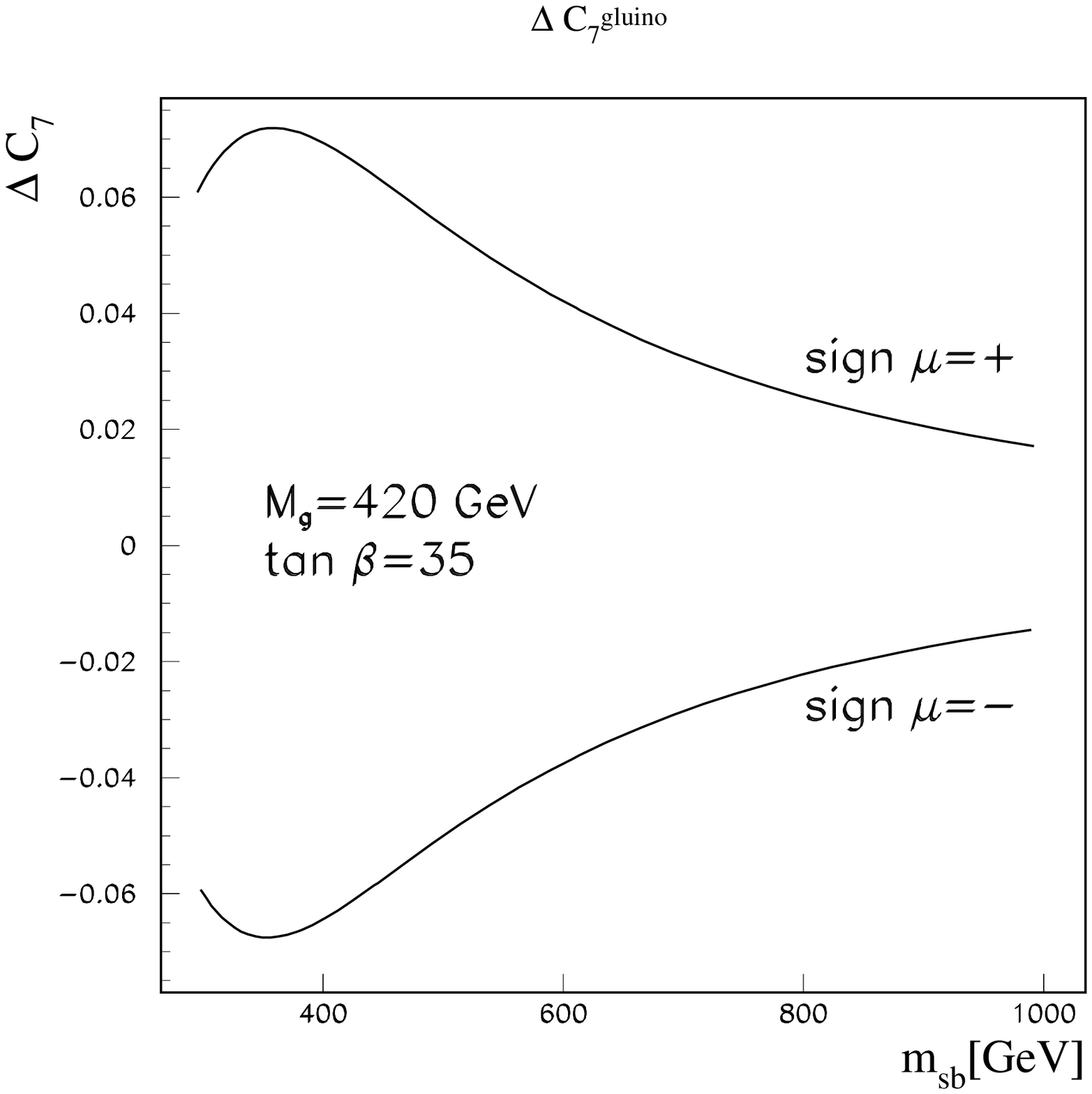} 
\hfill
\epsfxsize = 0.5\textwidth \epsffile{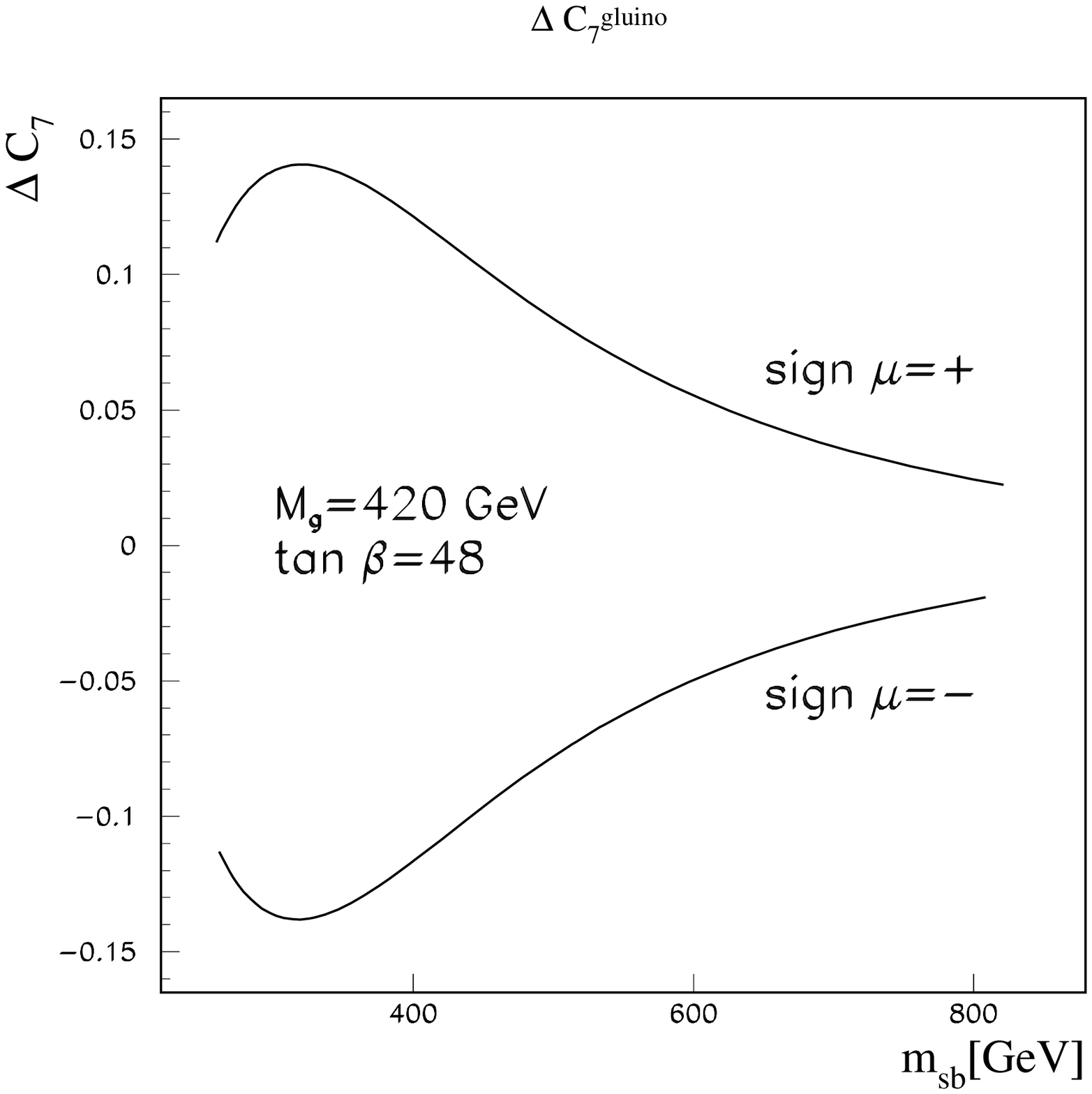}
}
\caption{\it 
Gluino mediated contribution to the Wilson coefficient $C_7$
as a function of the sbottom mass for fixed gluino mass 
$M_{\tilde g}=420$ GeV and $\tan\beta=35,$ 48 as indicated in
figures. The sign of the contribution depends on sign$(\mu).$
\vspace*{0.5cm}}
\label{fig:c7}
\end{figure}

The decay width of $b \to s \gamma$ can be written 
\begin{equation}
 \Gamma(b \to s \gamma) = 
 \frac{m_b^5 \, G_F^2 \, |V_{tb} V_{ts}^*|^2
 \, \alpha}{32 \pi^4} 
\left\vert C^{eff}_7  \right\vert^2  \,,
\end{equation}
where in our model
\bea
\left\vert C^{eff}_7  \right\vert^2 = 
\left\vert C_7 + C^{\tilde g}_7 \right\vert^2 +
\left\vert C^{'\tilde g}_7  \right\vert^2 \,. 
\eea
Here 
\begin{eqnarray}
C^{\tilde g}_7 & = &  
 - \frac{16 \sqrt{2} \pi^3 \alpha_s(\mu_b)} {G_F \, V_{tb}V^*_{ts}} 
\left[           C_{7b,\tilde{g}}(\mu_b) + 
 \frac{1}{m_b}   C_{7\tilde{g},\tilde{g}}(\mu_b)
\right] \,,
 \nonumber  \\ 
C^{'\tilde g}_7 & = &
 - \frac{16 \sqrt{2} \pi^3 \alpha_s(\mu_b)} {G_F \, V_{tb}V^*_{ts}} 
\left[           C'_{7b,\tilde{g}}(\mu_b) + 
 \frac{1}{m_b}   C'_{7\tilde{g},\tilde{g}}(\mu_b)
\right] \,,
\label{c7hat}
\end{eqnarray}
and $C_7$ stands for the total contribution from the effective
Hamiltonian \Eq{weffham}. In the SM its value is $C_7^{SM}=-0.3$ 
\cite{buras}. In the MSSM $C_7$ receives large contribution 
from charged Higgs loops which add constructively with $C_7^{SM}$ and
at large $\tb$ also from chargino loops which add constructively
(destructively) with $C_7^{SM}$ for $sign(\mu)=-1$ ($sign(\mu)=+1$).
Therefore for large $\tb$ and small Higgs and sparticle masses 
the $C_7$ may be completely dominated by SUSY.
Therefore the present experimental result \cite{cleo}
\begin{equation}
    2 \times 10^{-4} 
  <  {\rm BR}(\bar{B}\to X_s\gamma)
  < 4.5\times 10^{-4},  
\label{cleoband}
\end{equation}
which favours the following allowed range:
\bea
 0.25 < |C^{eff}_7| < 0.375 \,,
\label{c7range}
\eea
implies strong constraints on the SUSY masses.

Let us now study the gluino contribution to $C^{eff}_7$ numerically.
In Fig. \ref{fig:c7} we plot the values of $C^{\tilde g}_7$ as a function
of the lightest bottom squark mass $m_{\tilde b_1}$ for a fixed
value of the gluino mass $M_{\tilde g}=420$ GeV and for two values of 
$\tb$ and $sign(\mu),$ as indicated in the figure. Note that this is 
the smallest gluino mass consistent with chargino mass bounds in our model. 
The values of $C^{\tilde g}_7$ might be sizable for small sparticle masses
but decreases rapidly if the masses are higher. The important
behaviour to notice is that the sign of  $C^{\tilde g}_7$ depends 
on the  $sign(\mu)$ exactly the same way as the dominant contribution
$C_7.$ This implies that $C_7$ and $C^{\tilde g}_7$ add up constructively
and no cancellation between them is possible. 
It has been argued in Ref. \cite{bghw,cmw} that the  
constraints on the SUSY parameter space coming from  $C_7$ 
which are rather restrictive  can be relaxed by new flavour 
contributions. However, in our model this does not happen and 
the $b\to s \gamma$ bounds  cannot be relaxed with help 
of $C^{\tilde g}_7.$ 

The gluino contribution $C^{\tilde g}_7$ is induced by the mixings 
in the left-handed down squark sector, thus by $V_d\equiv V_{CKM}.$
However, the   coefficient  $C^{'\tilde g}_7$ is almost entirely
induced by the mixings in the right-handed down squark sector.
If these mixings are similar in size to the CKM mixings, as we 
argue here, then in our scenario with large $\tb$,  $C^{'\tilde g}_7$
is as large as   $C^{\tilde g}_7.$ While its absolute contribution
to $C^{eff}_7$ is subdominant (see Fig. \ref{fig:c7}) it still may
have important phenomenological consequences. Namely,
 while the mixing induced time dependent CP asymmetry in $B\to M_s \gamma$,
where $M_s$ is some CP eigenstate, is predicted to be very small,
a few percent,  in the SM;  our model, where it can be expressed as
\bea
\frac{\Gamma(t)-\bar \Gamma(t)}{\Gamma(t)+\bar \Gamma(t)}=
A_t \sin\Delta M_{B_d}t\,, \;\;\;\;\;
A_t=\frac{2 \mrm{Im}\left[   
e^{-i\theta_{B_d}} ( C_7 + C^{\tilde g}_7) C^{'\tilde g}_7
\right]}{|C^{eff}_7|^2} \,,
\eea
with $\theta_{B_d}=arg(M_{12}^{B_d})$ being the phase in the $B-\bar B$
mixing amplitude, asymmetries $A_t$ of more than 10\%
(which would be a clear and powerful signal of beyond the Sm physics)
are allowed.  Similar conclusions hold also in models with 
right-handed neutrinos \cite{bgoo}.

\subsection{Lepton Flavour Violation}

So far we have shown that in SUSY SU(5) the new flavour physics 
contribution to flavour changing hadronic observables is subdominant. 
At the same time the SUSY contribution to $b\to s\gamma$ induced by 
the effective Hamiltonian \Eq{weffham} constrains severely the SUSY
scale in our model. 
Now we turn to study the LFV processes which are completely dominated by
the new physics. 
The amplitude for the process $l_j\to l_i\gamma^*$ where the photon
is off-shell can be written as
\bea
M = e \epsilon^{\alpha *}\bar l_i \left[ 
q^2 \gamma_\alpha \left(A_1^L P_L + A_1^R P_R \right) + 
m_{l_j}i\sigma_{\alpha\beta}q^\beta\left(A_2^L P_L + A_2^R P_R \right)
\right] l_j  \,,
\eea
where $q$ is the photon momentum and $A_{1,2}^{L,R}$ are the form factors
giving rise to the process. Note that all the form factors contribute
to $\mu-e$ conversion but only $A_{2}^{L,R}$ give rise to the lepton
radiative decays. The form factors are induced by two types of loop
diagrams with neutralinos and charged sleptons in the loop, and 
charginos and sneutrinos in the loop:
\bea
A_{1,2}^{L,R}=A_{1,2}^{(n)L,R}+A_{1,2}^{(c)L,R}\,,
\eea
where schematically written
\bea
A_{1,2}^{(n)L,R} &=& A_{1,2}^{(n)L,R}
\left(M_{\tilde\chi^0}, m_{\tilde l}, N, \Gamma_E \right)\,, \nn\\
A_{1,2}^{(c)L,R} &=& A_{1,2}^{(c)L,R}
\left(M_{\tilde\chi^+}, m_{\tilde\nu}, O_{L,R}, \Gamma_\nu \right) \,,
\eea
depend on the slepton masses and mixings as well as on the neutralino and 
chargino masses and mixings. We have adopted the formulas for the form 
factors from Ref. \cite{japsN} and we do not present them here.
The decay rate of $l_j\to l_i\gamma$ is then given by
\bea
\Gamma(l_j\to l_i\gamma)=\frac{e^2}{16\pi}m_{l_j}^5 
\left( |A_{2}^L|^2 +  |A_{2}^R|^2  \right) \,.
\eea

\begin{figure}[t]
\centerline{
\epsfxsize = 0.5\textwidth \epsffile{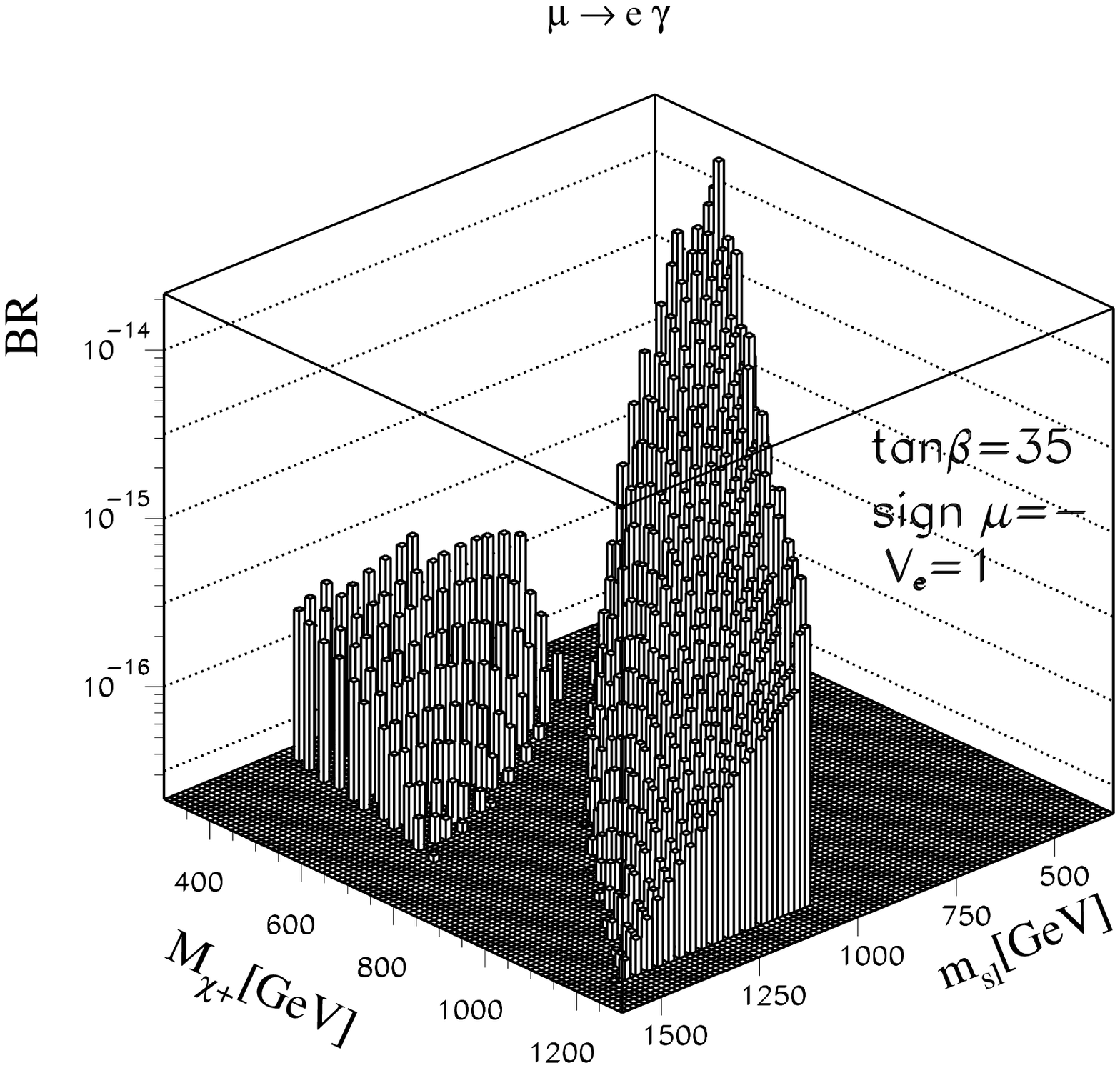} 
\hfill
\epsfxsize = 0.5\textwidth \epsffile{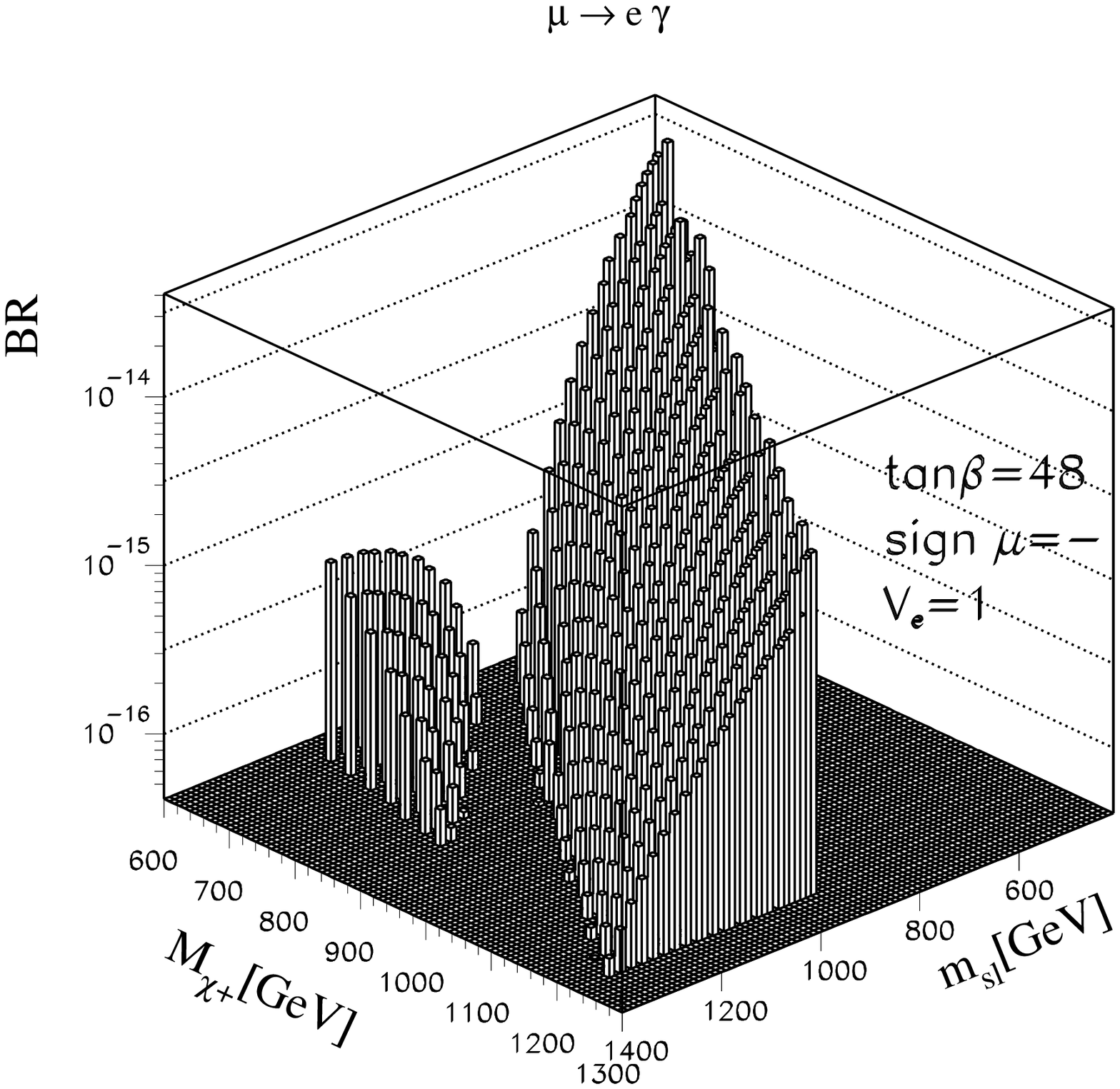}
}
\caption{\it 
The branching ratio of $\mu\to e\gamma$ as a function of the 
lightest charged slepton and the lightest chargino mass 
assuming that the flavour mixing in the left-slepton sector is
vanishing, $V_e=\mrm{\bf 1}.$ We have fixed sign$(\mu)=-$ and   
values of $tan\beta$ are indicated in the figures. 
The destructive interference between the gaugino and higgsino
contributions suppresses the branching ratio in large regions 
of the parameter space.  The constraints 
 coming from $b\to s\gamma$ are taken into account.
\vspace*{0.5cm}}
\label{fig:muejap}
\end{figure}

We start with studying the decay $\mu\to e \gamma$ in the case
where 
the only source of LFV is the mixing in the right-handed slepton sector
as given by \Eq{vgut}, and the mixing matrix $V_e$ is the unit matrix. 
This is the case studied in Ref. \cite{japs}.
In Fig. \ref{fig:muejap} we plot the branching ratio of 
$\mu\to e\gamma$ on the plane of the lightest charged slepton mass
$m_{\tilde l_1}$ and the lightest chargino mass $M_{\tilde \chi^+_1}$
for two values of $\tb=35,\,48$ and $sign(\mu)=-1$ 
(to achieve Yukawa unification).
We have taken into account the experimental constraints 
coming from $b\to s\gamma$ by requiring that the total value of $C^{eff}_7$
is in the allowed range \rfn{c7range}.  
As seen in Fig. \ref{fig:muejap} the constraint \Eq{c7range} puts
strong lower bounds on the lightest sparticle masses.
Our values of the branching ratio of $\mu\to e \gamma$
are much smaller than the quoted values in Ref. \cite{japs}.
The reason is twofold. First, the $b\to s\gamma$ constraint pushes
the sparticle masses to high values\footnote{It has been noticed in 
Ref. \cite{b2} that for a particular corner of the parameter space where
$M_0\ll m_0\sim A$ one can supress $b\to s\gamma$ and 
still have light gauginos. However, also LFV processes are suppressed 
for this parameter space.} and this has not been taken
into account in Ref. \cite{japs}. Second,  the authors of Ref. \cite{japs}
fix the top Yukawa coupling at $M_{GUT}$ to be $f_t^G=1.4$ which
for large $\tb$ implies by far a too large top quark mass.
For $m_t=174$ GeV and  $\tb=35$, the correct value is $f_t^G=0.56.$
For large part of the parameter space the  
destructive interference between the gaugino and higgsino
contributions suppresses the $\mu\to e \gamma$ branching ratio
to almost vanishing values.

\begin{figure}[t]
\centerline{
\epsfxsize = 0.5\textwidth \epsffile{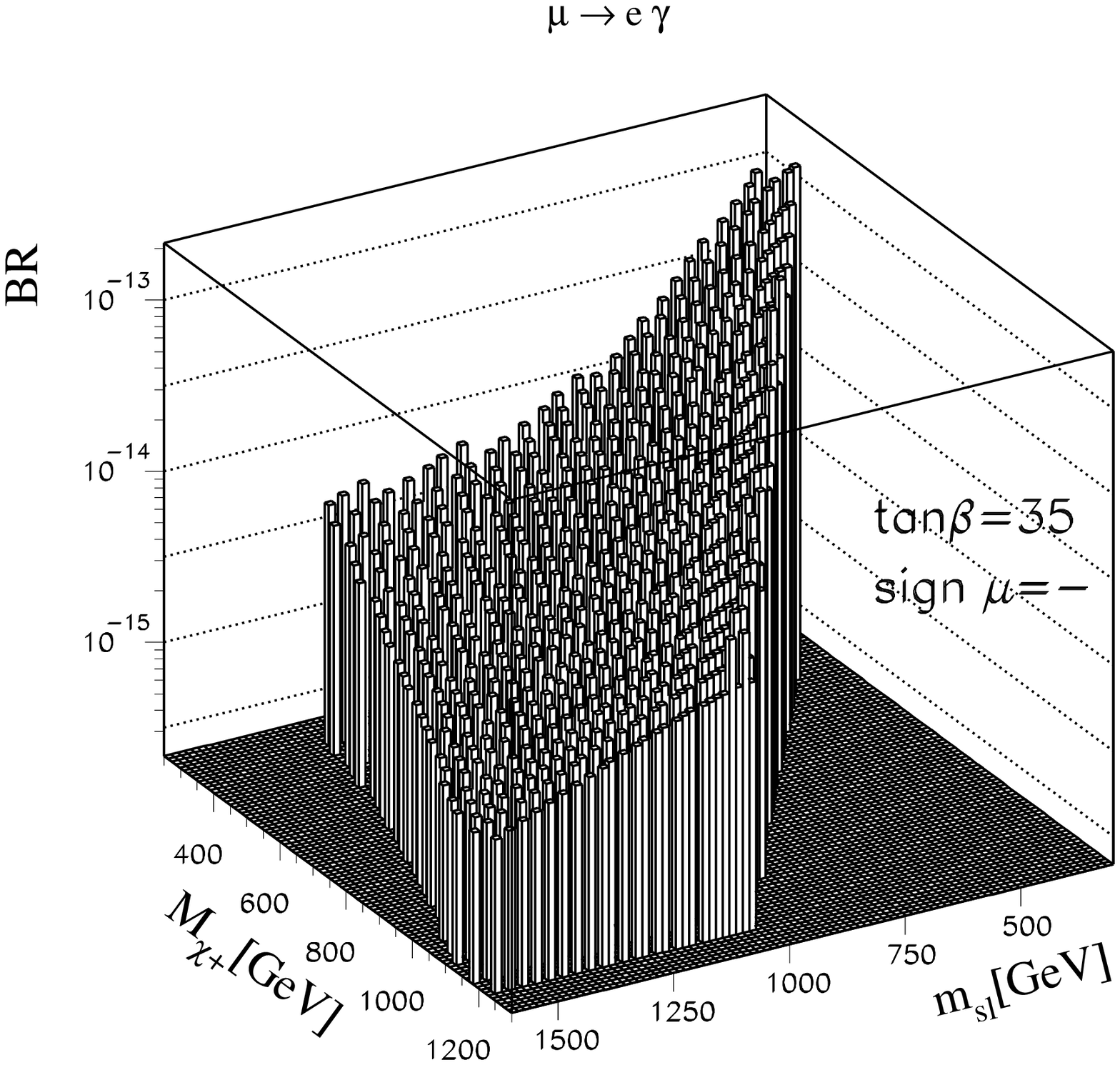} 
\hfill
\epsfxsize = 0.5\textwidth \epsffile{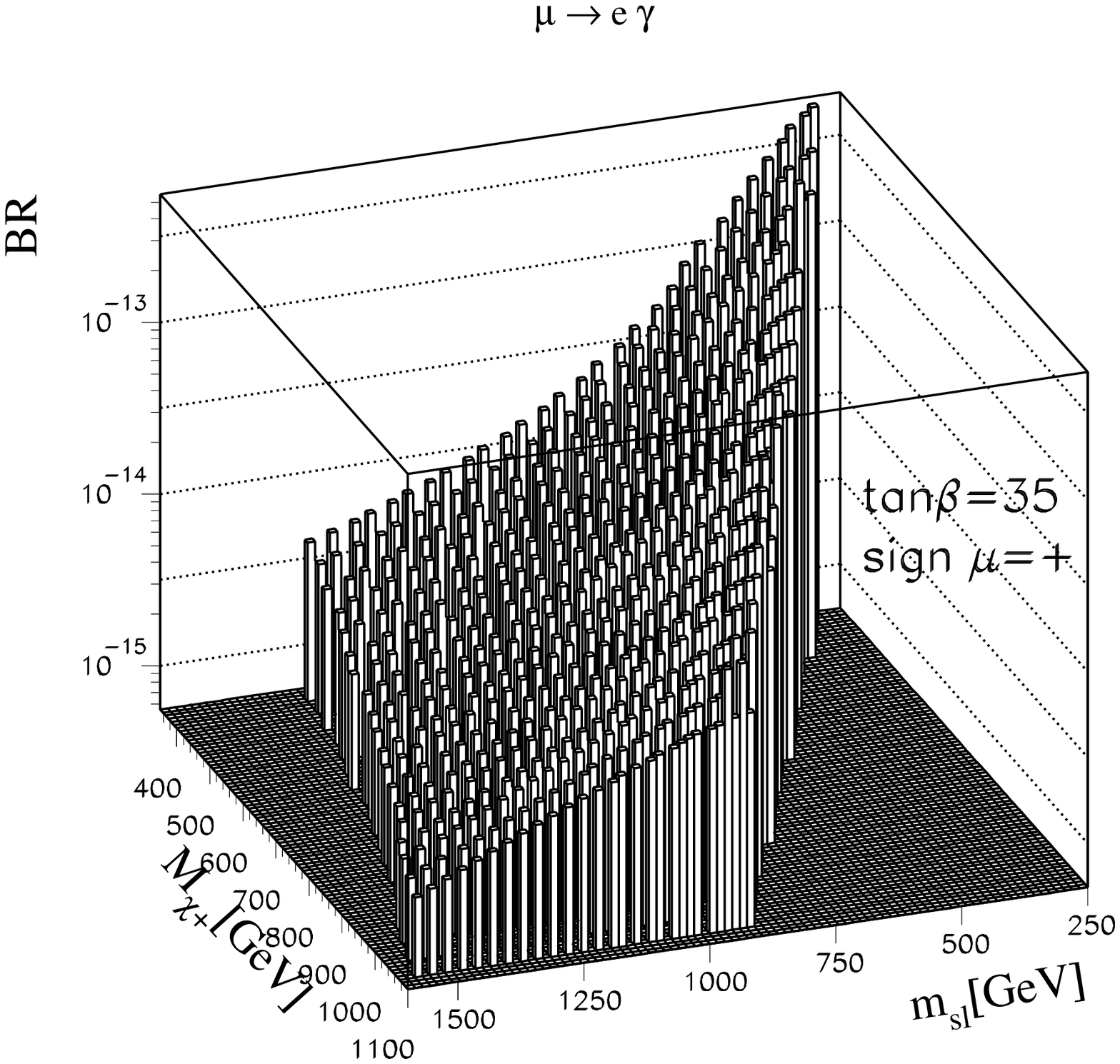}
}
\caption{\it 
The branching ratio of $\mu\to e\gamma$ as a function of the 
lightest charged slepton and the lightest chargino mass for
fixed $tan\beta=35.$ The sign of the $\mu$ parameter is 
indicated in the figures. Here and in the following 
$V_e^{ij}=U_e^{ij}=(V_{CKM}^G)^{ij}.$ 
The constraints on SUSY mass spectrum
 coming from $b\to s\gamma$ are taken into account.
\vspace*{0.5cm}}
\label{fig:mue35}
\end{figure}

\begin{figure}[t]
\centerline{
\epsfxsize = 0.5\textwidth \epsffile{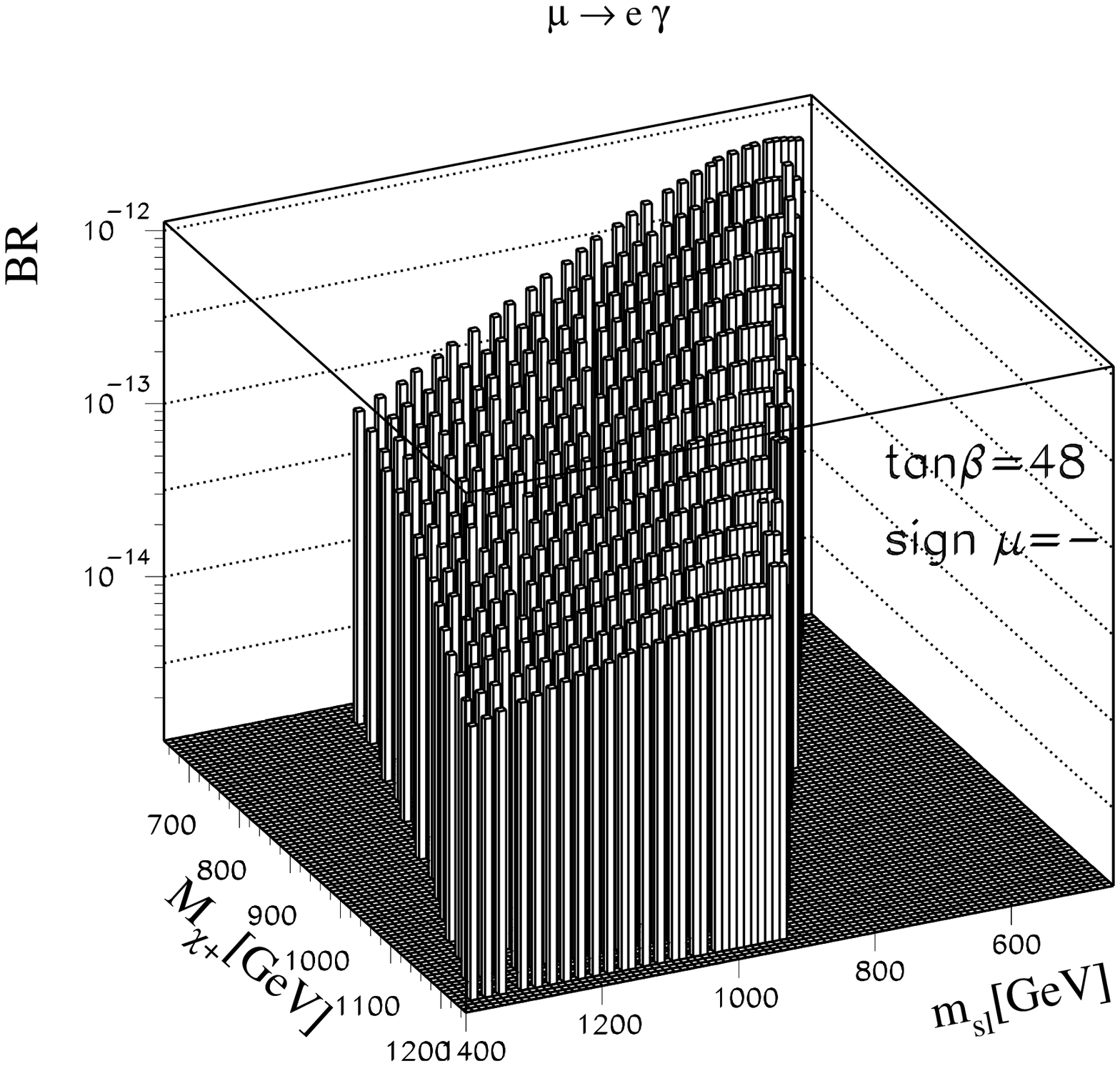} 
\hfill
\epsfxsize = 0.5\textwidth \epsffile{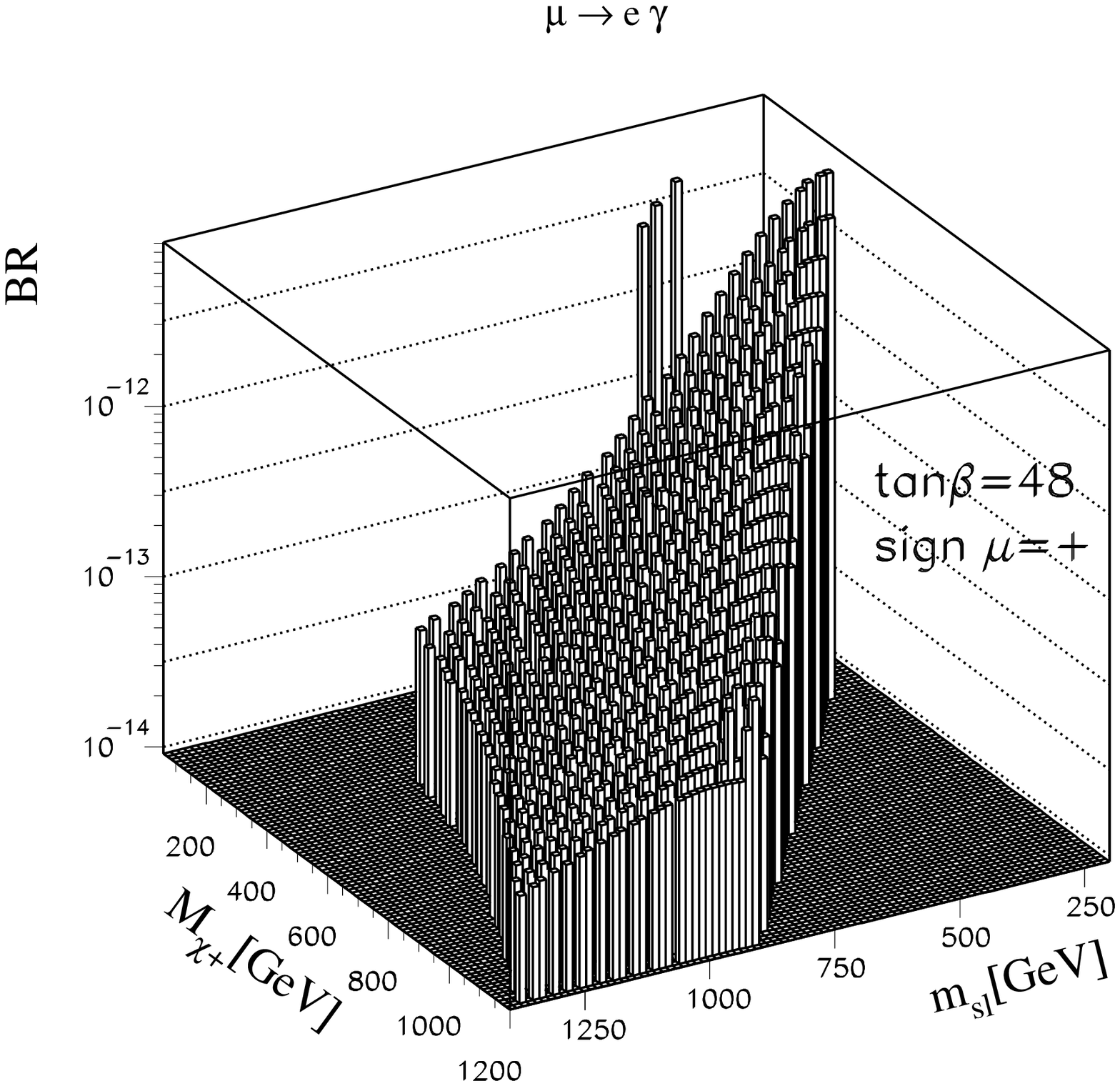}
}
\caption{\it 
The same as in Fig. \ref{fig:mue35} but for $tan\beta=48.$ 
\vspace*{0.5cm}}
\label{fig:mue48}
\end{figure}

The situation changes completely if we allow also flavour mixings 
in the left-slepton sector and allow $sign(\mu)$ to be also positive.
Taking $V_e^{ij}=U_e^{ij}=(V_{CKM}^G)^{ij}$ as we predict in our model
we plot the $\mu\to e \gamma$ branching ratio for $\tb=35$ in
Fig. \ref{fig:mue35} and for $\tb=48$ in Fig. \ref{fig:mue48}.  
The branching ratios are about two orders of magnitude higher than
in Fig. \ref{fig:muejap} and no cancellation occurs for any
sparticle masses. This implies that $\mu\to e \gamma$ is dominated
by the sneutrino-chargino contribution. The LFV pattern here is
exactly the same as in models with right-handed neutrinos.
Notice that for  $sign(\mu)=+1$ the sparticle masses are allowed to 
be much smaller than in the other case. This is because for 
$sign(\mu)=+1$ the chargino contribution to  $b\to s\gamma$
interferes destructively with the SM and charged Higgs contributions.
In particular, for $\tb=48$ and for very small chargino and slepton masses
the chargino contribution can be so large that it cancels the SM
and $H^+$ contributions and gives the allowed $C_7^{eff}$ value with 
an opposite sign. This is seen   in Fig. \ref{fig:mue48} for 
$sign(\mu)=+1$ in which a small parameter region around
$M_{\tilde \chi^+_1}\approx 100$ GeV and $m_{\tilde l_1}\approx 300$ GeV
is allowed. This region can be excluded by collider searches 
for a very light chargino or by improving the bound on the 
$\mu\to e \gamma$ branching ratio by a factor of few.


\begin{figure}[t]
\centerline{
\epsfxsize = 0.5\textwidth \epsffile{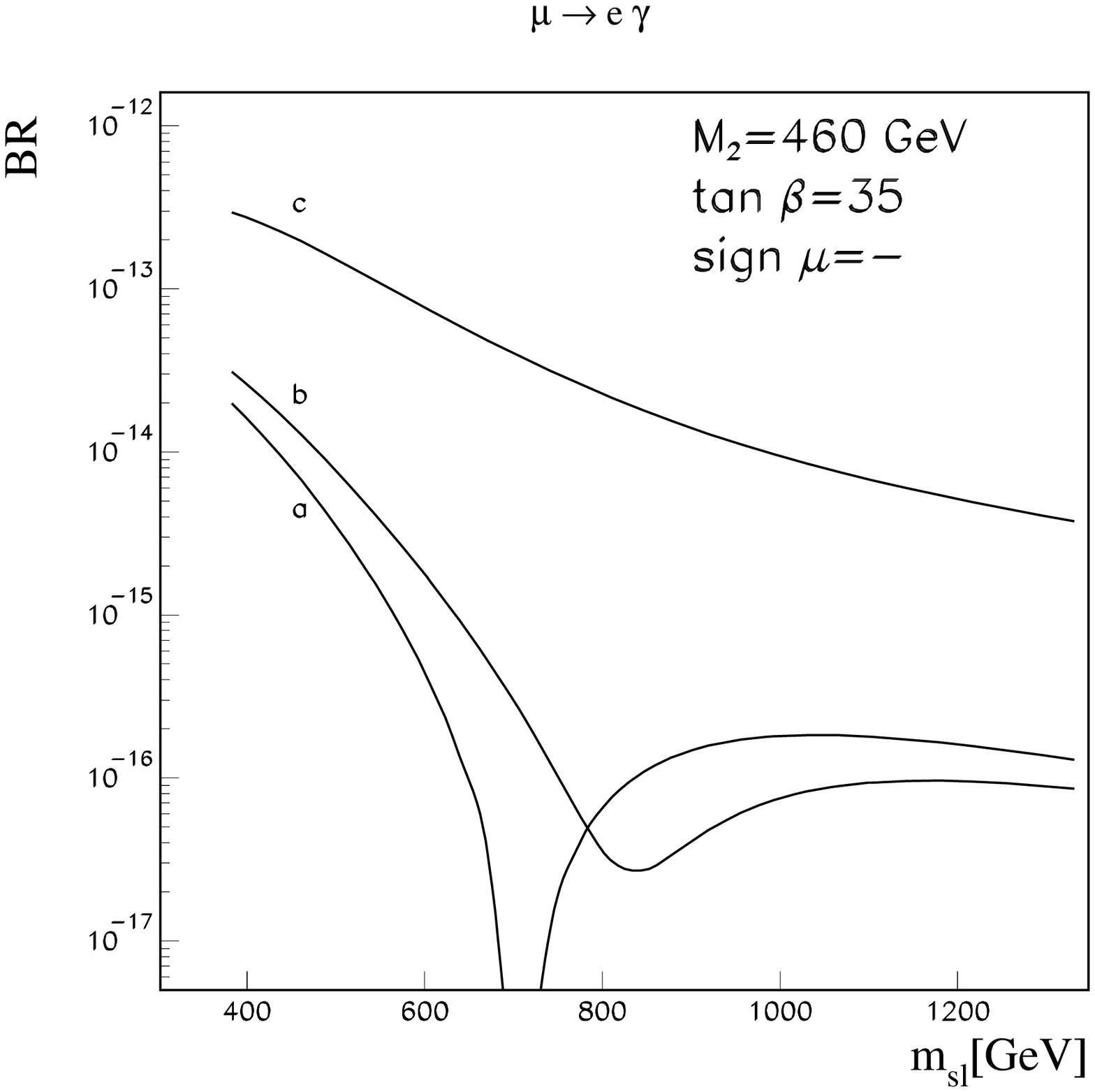} 
\hfill
\epsfxsize = 0.5\textwidth \epsffile{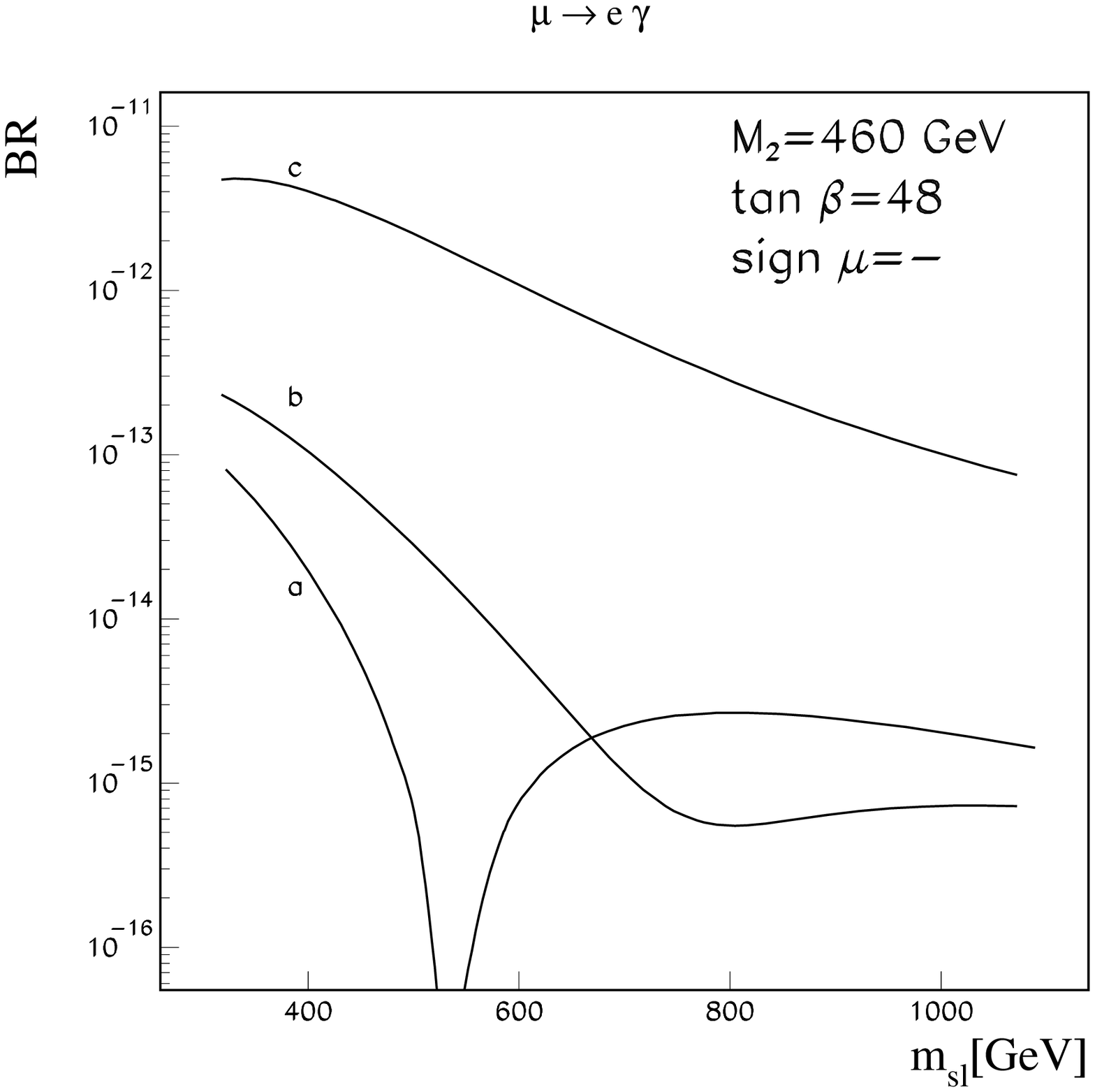}
}
\caption{\it 
The branching ratio of $\mu\to e\gamma$ as a function of the 
lightest charged slepton mass for fixed wino mass $M_2=460$ GeV.
We have fixed sign$(\mu)=-$ and values of $tan\beta$ are indicated 
in the figures. For the curves denoted by $(a)$ $V_e^{ij}=0,$
$i\neq j,$ thus they correspond to Fig. \ref{fig:muejap}.  
For the curves $(b)$ $V_e^{ij}=0.1\times U_e^{ij}$ and for the 
curves $(c)$ $V_e^{ij}=U_e^{ij}.$ 
\vspace*{0.5cm}}
\label{fig:muem}
\end{figure}

\newpage
We have shown that the non-vanishing flavour mixings in the left-slepton
sector are crucial to ensure the detectibility of $\mu\to e \gamma$
in the planned experiments. Since there is no experimental information
on these mixings the central question to ask now is how small the
off-diagonal elements of $V_e$ can be and still allow successful determination
of  $\mu\to e \gamma.$ To analyze this question we plot
in Fig. \ref{fig:muem} the $\mu\to e \gamma$ branching ratio against
the lightest slepton mass $m_{\tilde l_1}$ for a fixed $M_{2}=460$ GeV
which is roughly the minimal chargino mass for $sign(\mu)=-1.$ 
The curves denoted by $a$ correspond to $V_e=\mrm{\bf 1},$ 
curves denoted by $b$  to $V_e^{ij}=0.1\times U_e^{ij},$ $i\neq j$ and
curves denoted by $c$  to $V_e^{ij}= U_e^{ij}.$ 
As can be seen, if the off diagonal elements of $V_e$ are as small as
10\% of the corresponding $U_e$ elements then the deep
cancellation is superseeded. Let us also mention that for the chosen
chargino mass  in Fig. \ref{fig:muem} the SUSY scale, 
$M_{SUSY}=\sqrt{m_{\tilde t_1}m_{\tilde t_2}},$ is
 $M_{SUSY}\approx 1200$ GeV for the minimally allowed slepton mass.
Thus the LFV processes are sensitive to SUSY scale above TeV.

Our calculations show that the rate of $\mu-e$ conversion 
in nuclei is about $6\times 10^{-3}$ times the branching ratio of 
$\mu\to e\gamma.$ The qualitative behaviour of the $\mu-e$ conversion
rate with the sparticle masses is the same as in the case of 
$\mu\to e\gamma.$ Thus the results for  $\mu-e$ conversion can be 
obtained by rescaling the figures for $\mu\to e\gamma.$
Therefore we do not present any new plots for the $\mu-e$ conversion
process. We conclude that the planned $\mu-e$ conversion 
experiments \cite{meco} are as sensitive to our models as the 
planned $\mu\to e\gamma$ experiments \cite{mueprop}.

\begin{figure}[t]
\centerline{
\epsfxsize = 0.5\textwidth \epsffile{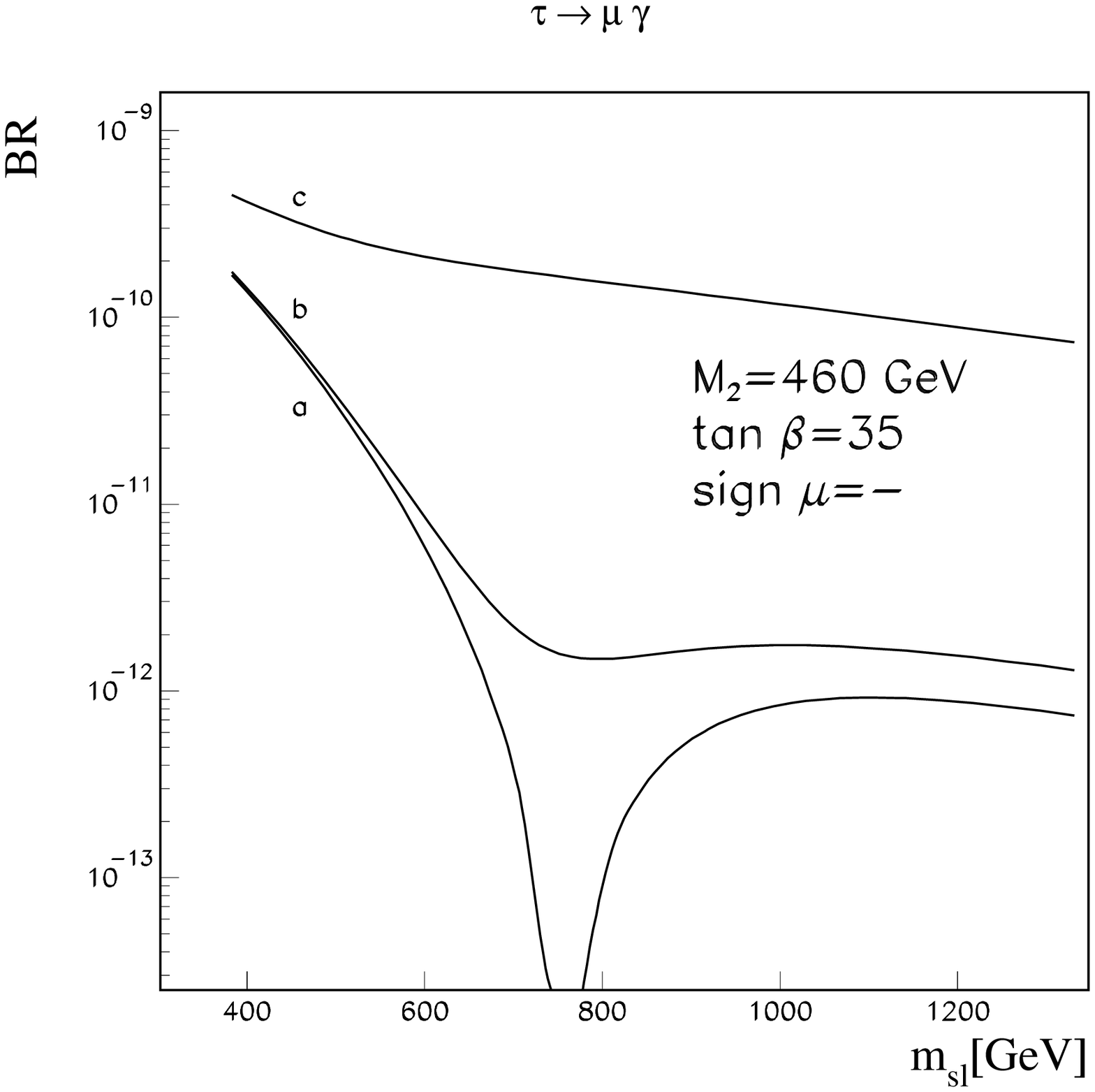} 
\hfill
\epsfxsize = 0.5\textwidth \epsffile{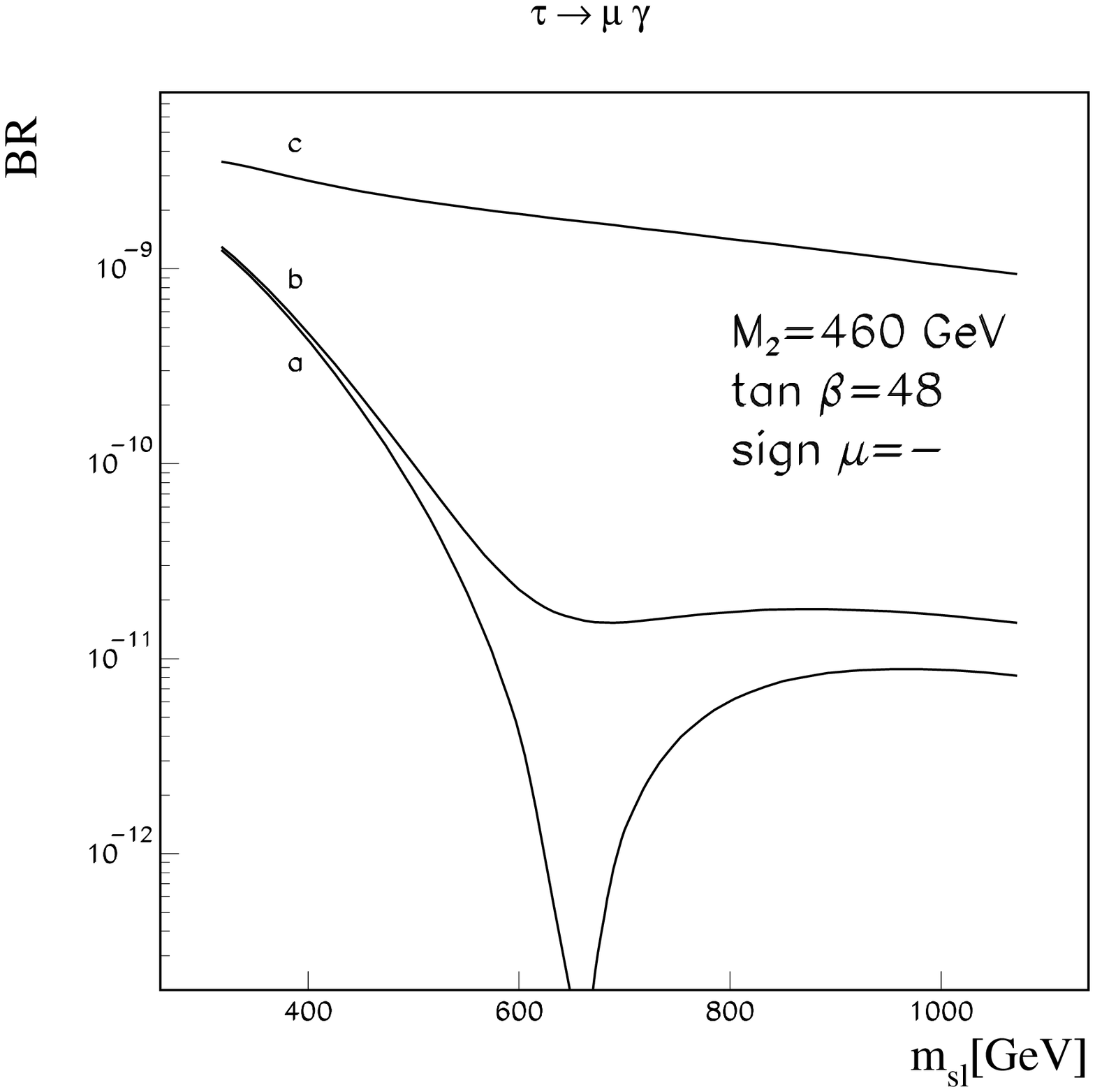}
}
\caption{\it 
The same as in Fig. \ref{fig:muem} but for the decay $\tau\to\mu\gamma.$
\vspace*{0.5cm}}
\label{fig:mutaum}
\end{figure}

Finally let us discuss the decay  $\tau\to \mu\gamma.$
In Fig. \ref{fig:mutaum} we plot the branching ratio of the 
decay  $\tau\to \mu\gamma$ for the same choice of parameters 
as in  Fig. \ref{fig:muem}. Even for $\tb=48$  
the  branching ratio is always below a few times $10^{-9}$
and  unobservable in the planned experiments. 
Thus if $\tau\to \mu\gamma$ will be discovered at these experiments
this implies some other LFV scenario than 
the one considered in this work.

\section{Conclusions}

Motivated by the sensitivity of the running or approved experiments
to flavour violating processes we have studied flavour violation
in the minimal SUSY SU(5) GUT at large $\tb.$ 
In this case the flavour mixing occurs both in the left and right
slepton and squark mass matrices and are enhanced by the large value of
$\tb.$  We have calculated the new physics contributions to 
$K-\bar K$ and $B-\bar B$ mixings and  to the decays $b\to s\gamma,$
$\mu\to e\gamma,$ $\tau\to \mu\gamma$ and to $\mu-e$ conversion in nuclei.
To predict reliably the rates of these processes 
we have correctly taken into account the measured values of
the low energy parameters as well as the constraints on the SUSY
particle masses.

We found that in our model the new physics contributions to $\Delta M_K$ 
and $\Delta M_B$ are negligible, but might reach a 
 10\% level in $\epsilon_K$
if there exist new GUT phases. No useful constraints on the model parameters 
can be derived from these processes.

The decay $b\to s\gamma$ receives contributions from two sources of
flavour violation: from the loops proportional to the CKM matrix elements
and from the loops exhibiting new flavour violation in the squark
mass matrices. The latter contribution interferes constructively with the 
dominant chargino contribution. At large $\tb$ the experimental 
constraints on the $b\to s\gamma$ branching ratio imply stringent
constraints on the SUSY particle masses, especially for $sign(\mu)=-1$
as required by Yukawa unification. In this case, the SUSY scale is 
constrained to be at least TeV. For such a  high squark masses the new
flavour physics contribution to  $b\to s\gamma$ branching ratio 
is a few percent. Nevertheless, this may induce CP asymmetries
considerably larger than in the SM.

There is a competition between the sensitivity of the future LFV experiments 
to the new flavour physics and the constraints on the SUSY scale 
coming from the $b\to s\gamma$ branching ratio. 
If the branching ratio of the decay $\mu\to e\gamma$ will be tested 
down to $10^{-14}$ and the SUSY scale is below 1 TeV then,   
the present scenario predicts that $\mu\to e\gamma$ should be discovered 
in these experiments. The branching ratio of the decay $\tau\to\mu\gamma$
is, however, predicted to be below a few times $10^{-9}$ and
should not be seen at LHC in the minimal SUSY SU(5).

\vspace{2.5cm}

\begin{center}
{\bf Acknowledgements}
\end{center}
We are very grateful to K. Jakobs and 
E. Ma for discussions and to J. Hisano for
clarifying communication.
The work of GB is supported by the DFG, 
that of KH is partially supported by the Academy of Finland
project no. 163394 and the one of MR by U.S. Department of Energy
under Grant No. DE-FG03-94ER40837.

\end{document}